\documentclass[useAMS,usenatbib]{mn2e}

\usepackage[T1]{fontenc}
\usepackage{aecompl}

\usepackage{graphicx}

\usepackage{rotating}
\usepackage{amsmath}

\usepackage{float}
\usepackage{lscape}
\usepackage{longtable}
\usepackage{supertabular}

\usepackage{aas_macros}

\def\s{\,\mathrm{s}}

\usepackage{amssymb}

\def\apjs{ApJS}

\begin{document}
\title{On the nature of the ``hostless'' short GRBs}
\author[R.~L.~Tunnicliffe et al.]
{\parbox{\textwidth}{R.L.~Tunnicliffe,$^{1}$
A.~J.~Levan,$^{1}$
N.~R.~Tanvir,$^{2}$
A.~Rowlinson,$^{3}$
D.~A.~Perley,$^{4}$ \\
J.~S.~Bloom,$^{4}$
S.~B.~Cenko,$^{4}$
P.~T.~O'Brien,$^{2}$
B.~E.~Cobb,$^{4}$
K.~Wiersema,$^{2}$
D.~Malesani,$^{5}$
A.~de Ugarte Postigo,$^{5}$
J.~Hjorth,$^{5}$
J.~P.~U.~Fynbo$^{5}$ and
P.~Jakobsson$^{6}$}
\vspace{0.4cm}\\
\parbox{\textwidth}{$^{1}$Department of Physics, University of Warwick, Coventry, CV4 7AL, UK\\
$^{2}$Department of Physics and Astronomy, University of Leicester, Leicester, LE1 7RH, UK\\
$^{3}$Astronomical Institute, University of Amsterdam, Science Park 904, 1098 XH Amsterdam, The Netherlands\\
$^{4}$Department of Astronomy, University of California, Berkeley, CA 94720-3411, USA\\
$^{5}$Dark Cosmology Centre, Niels Bohr Institute, University of Copenhagen, 2100 Copenhagen, Denmark\\
$^{6}$Centre for Astrophysics and Cosmology, Science Institute, University of Iceland, Dunhaga 5, IS-107 Reykjavik, Iceland}}

\date{Received;Accepted}
\maketitle

\begin{abstract}
A significant proportion ($\sim30\%$) of the short-duration gamma-ray bursts (SGRBs) localised by {\em Swift} have no detected
 host galaxy coincident with the burst location to deep limits, and also no high-likelihood association with proximate
 galaxies on the sky. These SGRBs may represent a population at moderately high redshifts ($z\gtrsim1$),
 for which the hosts are faint, or a population where the progenitor has been kicked far from its host or is sited
 in an outlying globular cluster. We consider the afterglow and host observations of three ``hostless'' bursts
 (GRBs 090305A, 091109B and 111020A), coupled with a new observational diagnostic to aid the association of SGRBs
 with putative host galaxies to investigate this issue. Considering the well localised SGRB sample,
 7/25 SGRBs can be classified as ``hostless'' by our diagnostic. Statistically, however, the proximity of these
 seven SGRBs to nearby galaxies is higher than is seen for random positions on the sky.
 This suggests that the majority of ``hostless'' SGRBs have likely been kicked from proximate galaxies at moderate redshift.
 Though this result still suggests only a small proportion of SGRBs will be within the AdLIGO horizon for NS-NS
 or NS-BH inspiral detection ($z\sim0.1$), in the particular case of GRB\,111020A a plausible host candidate is at $z=0.02$.
\end{abstract}

\section{Introduction}

Short-duration gamma-ray bursts (SGRBs)\footnote{Usually taken as those with
 $T_{90}<2$\,s, where $T_{90}$ is the time over which 90\% of the prompt gamma-ray
 emission is observed.  As a class, the SGRBs also have harder spectra
 than LGRBs, but there is significant scatter.} reside in different environments from 
 their long GRB counterparts (LGRBs). While, LGRBs are usually found
 in the brightest regions of star forming host
 galaxies \citep{GRBoffset,LGRBs_in_bright_regions}, SGRBs 
 often appear in galaxies with established 
 older populations with a handful linked strongly to elliptical galaxy hosts 
 \citep{Gehrels_050509B_host,Bloom_2006_050509B,Berger_2005_050724_host}. 
 This diversity in environments almost certainly reflects differing progenitors for the LGRBs and SGRBs \citep{Bloom_2006_diverse_progenitors}. 
 Indeed, in the majority of LGRBs at low redshift
 it has been possible to isolate the signature of a type Ic supernova,
 suggesting that their progenitors are Wolf-Rayet stars \citep[][and references therin]{SN_connection}. 
 In contrast, deep searches in SGRBs at similar redshift fail to locate any such 
 signatures, and offer further evidence that
 the progenitors of short bursts are not related to stellar core 
 collapse \citep{Bloom_2006_050509B,Hjorth_2005_050709_OA,Rowlinson_2010_080905A_Opt}. The varied host demographics, offer part of the picture,
 but the locations of the bursts on their hosts
 are also greatly diagnostic. It appears that short bursts are 
 scattered significantly on their hosts, occurring in typically
 fainter regions, and at larger offsets than their long
 cousins \citep{LGRBs_in_bright_regions,Fong_2010_SGRBhosts_HST,Church_NS_offsets}. These properties can naturally be explained
 if the progenitor is a merger of two compact objects (e.g.,
 neutron star -- neutron star (NS), collapsing to a black hole by accretion-induced collapse (AIC), or neutron star -- black hole (BH))
 \citep{mergingNSbinary_could_emit_gammarays,FryerNS_dist,spatialdistns_ns,Fong_2010_SGRBhosts_HST}. This 
 model has proved extremely hard to test observationally, although the recent discovery of a possible kilonova in
 the SGRB 130603B offers support for the model for at least some SGRBs \citep{tanvir13,berger13}. 

A key distinguishing factor between different intrinsically ancient progenitor populations, 
 such as accretion-induced collapse (AIC) of white dwarfs to neutron stars \citep{Other_SGRB_progenitors,Metzger_2006_Magnetars}, and compact binary 
 mergers, comes from the dynamics of the systems themselves.
 In a double compact object (DCO) binary, a combination of natal kicks, and mass loss from the 
 binary at the time of each supernova, can act to provide 
 the systems with space velocities
 of several hundred $\mathrm{km}\s^{-1}$ (e.g. \citealt{2010_8_ns_binaries}). Integrated over the lifetime of the 
 binary of $10^7-10^{10}$ years this can correspond to distances of tens of $\mathrm{kpc}$ from their birth sites, although the
 fraction of binaries that attain these distances remains uncertain \citep{Belczynski_2006_NSmerger,Church_NS_offsets}.
 For extremely high kicks, or relatively low mass host galaxies, the binary may 
 escape the galactic potential of its host altogether.
 Hence,
 a population of SGRBs in intergalactic space would offer strong 
 support for a binary merger model for their progenitors \citep{Bloom_2006_diverse_progenitors,Berger_2010_nohost}.

However, determining the offset of
 a burst from its host is non-trivial when there is no obvious parent galaxy coincident with or close to the burst location, i.e. 
 where there is no outstanding candidate with an extremely low probability of chance alignment.
Probabalistic methods based on the sky density of galaxies are often used to argue for a host association
 (e.g. \citealt{GRBoffset, Bloom_2007_060502B_putative_hosts, AL_random_associations, Berger_2010_nohost}). However, 
 the sky density is such that for that any random position on the sky is likely to be within a few arcseconds of a galaxy with $R<25$.
 In other words, in many cases we cannot \emph{strongly} identify the host galaxy (probability of $\lesssim1\%$), 
 which can lead to mis-identifications. 
 A second problem is that to the limits of our ground-based
 (or even {\em Hubble Space Telescope}) observations, we probe 
 a reducing fraction of the galaxy luminosity function as we move to higher redshift. 
 Hence there is a potential for confusion between GRBs which have been kicked far, and hence are well offset, 
 from relatively local hosts and those which lie within fainter 
 galaxies at high redshifts. This problem is particularly acute if the high redshift galaxies host primarily old stellar
 populations, and hence exhibit only weak rest frame UV (observer-frame
 optical) emission.


In this sense the ``hostless" problem for SGRBs in not that there are a lack of candidate hosts;
 in all lines of sight there will be plausible parent galaxies within a few tens of kpc in projection.
 Frequently the probability of chance alignment with at least one of these is small, and may be suggestive of kicks to the SGRB progenitors
 \citep{Berger_2010_nohost}. Instead, the problem in these hostless cases is the difficulty in determining uniquely the parent galaxy 
 (from e.g. several with similar probabilities, or underlying larger scale structure).
 This means we are unable to make full use of the diagnostic information contained in the offset distribution and
 the properties of the hosts for improving our knowledge of SGRB progenitors. 

Obtaining the redshift for the GRB using the afterglow would allow us to narrow our search to hosts within a small redshift range. 
 However, for short bursts the faintness of their afterglows \citep{Kann_2011_OAG_TypeI_vs_TypeII} means that redshifts are difficult to obtain in practice, and in nearly 
 all cases to-date redshifts for SGRBs have been inferred from their presumed host galaxy rather than from the burst itself
 (e.g. \citealt{Hjorth_2005_050709_OA,Berger_2005_050724_host,Rowlinson_2010_080905A_Opt,Fong_2011_100117A_host}). In fact, 
 even accounting for the faint continuum, in some cases the lack of any absorption features  
 also gives an indication that the burst is not in a dense interstellar medium (ISM) \citep{Piranomonte_2008_070707_Beta-OX,Berger_GCN_100117A_Gemini_spec}. 


 Based on the optically localised SGRB sample outlined in section \ref{SGRB_sample} with details in appendix \ref{SGRB_sample_table}, $\sim70\%$ 
 have apparently well-associated host galaxies, while the remainder are apparently hostless.
 This may offer evidence for kicks. 
 Here, we present the discovery and subsequent observations of the optical afterglows and host limits of
 a further two hostless bursts, GRB\,090305A (see also \citealt{Berger_2010_nohost} and \citealt{NicuesaGuelbenzu_2012_preprint_090305}), 
 GRB\,091109B and deep limits to a third, GRB\,111020A (see also \citealt{Fong_GCN_111020A_Gemini}). These GRBs are all unambiguously of the short-hard class, 
 with $T_{90}<0.5\s$ and prompt emission which is spectrally hard. Indeed, even given the recent work suggesting a 
 shorter dividing line between the short- and long-GRB populations \citep{Bromberg_2013_short_swift}, these bursts would
 remain in the short-hard class. 
 We consider the extent to which our current detection limits probe the galaxy luminosity function as a function
 of redshift, and what this implies for hostless GRBs more generally. 

These hostless GRBs could reside within relatively high redshift (but so far unseen) host galaxies or
 have travelled far from their low redshift ($z<1$) hosts, perhaps within the 
 intergalactic medium (IGM).
 Clues to their origins may come from studies of the most likely hosts amongst the nearby
 galaxies on the sky. 
 We present an alternative diagnostic tool developed by taking random positions and comparing them to the distribution of galaxies on the sky, thus reproducing 
 the sort of analysis performed when looking for a short GRB host. From this we directly determine the chance probability of association ($P_{\rm chance}$)
 and a radius within which we can confidently state a host association. Doing this avoids assumptions as to the functional form of the
number counts, and naturally encapsulates field to field variance and clustering, resulting in more robust assessments of the
chance probability than the traditional route of using number counts alone.  

 
\section{Observations and Analysis} \label{Obs_section}

\subsection{GRB\,090305A}

\subsubsection{Prompt and X-ray observations}

GRB\,090305A was detected by by the Burst Alert Telescope (BAT) instrument on \emph{Swift} \citep{Barthelmy_2005_BAT_instrument} in 2009 March 05 at
 05:19:51 UT \citep{090305_Swiftdetection}. The GRB had a duration of $T_{90}=0.4\pm0.1\s$ and a fluence 
 ($15-150\,\mathrm{keV}$) of $7.5\pm1.3\times10^{-8}\,\mathrm{erg}\,\mathrm{cm}^{-2}$ with the errors quoted at the 90\% confidence level 
 \citep{Krimm_GCN_090305A_BAT_refined}. For a GRB at $z=0.3$ (a typical redshift for an SGRB) extrapolating this fluence to the $1-1000\,\mathrm{keV}$ range gives an 
 isotropic equivalent energy of $E_{iso}=1.6\times10^{50}\mathrm{erg}$. 

\begin{figure}
\begin{center}
\includegraphics{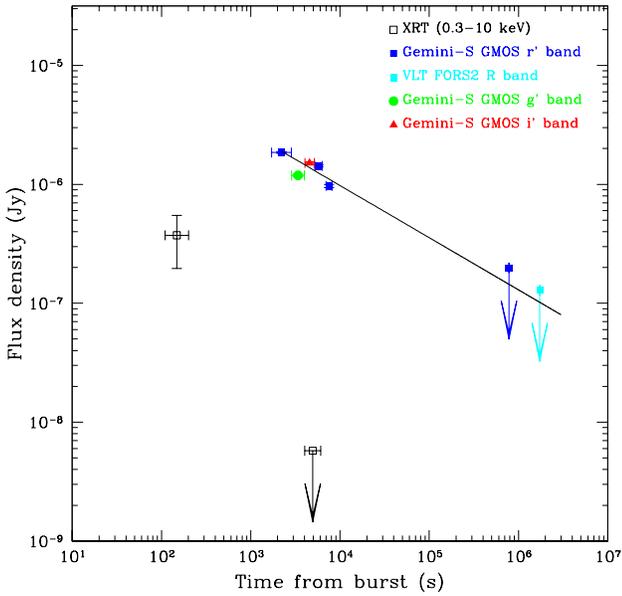}
\caption{X-ray and optical light curve of GRB\,090305A.
 The XRT measurements are shown as black open squares including the single XRT detection and late time upper limit \citep{Swift_product1}.
 The Gemini-S GMOS optical measurements and limit in the $r^{\prime}$, $g^{\prime}$ and $i^{\prime}$ bands are the blue filled squares,
 green circle and red triangle respectively. The VLT $R$ band limit is shown as a cyan square. Fitting a single power law to the $r^{\prime}$ band,
 where we have three detections and an upper limit, yields a decay index of $\alpha_r = 0.44\pm0.02$, shown on the plot. This is a poor fit to the data 
 but with only three detections it is difficult to constrain the slope.
 } \label{090305_X+O_lc}
\end{center}
\end{figure}

The X-ray Telescope (XRT) on \emph{Swift} \citep{Burrows_2005_XRT_instrument} began observations $103.4\s$ after the BAT trigger in photon counting (PC) mode 
 but no X-ray afterglow was detected \citep{090305_XRT_refined}. An optical afterglow was detected by the Gemini Multi-Object Spectrograph (GMOS)
 instrument on the Gemini-South telescope \citep{Hook_2004_GMOS_N_instrument} at position $\alpha=16^{\mathrm{h}}\,07^{\mathrm{m}}\,07.58^{\mathrm{s}}$,
 $\delta=-31^{\mathrm{o}}\,33\mathrm{'}\,22.1\mathrm{''}$ \citep{090305_GMOS}. 
 Using the position provided by the Gemini observations and by relaxing the default screening criteria, the X-ray afterglow
 was identified with 99.99\% confidence using the method of \cite{Confidence_lowcounts} for determining confidence limits with
 a low number of counts. The source was no longer detected when further XRT measurements were made $3.92\,\mathrm{ks}$ after the BAT trigger
 for $2.05\,\mathrm{ks}$. Using all available data this gives a $3\sigma$ upper limit of $1.7\times10^{-3}\mathrm{count}\,\s^{-1}$
 indicating a decay slope of at least $\sim0.8$ \citep{090305_XRT_refined}.
 All BAT and XRT measurements and limits are shown in figure \ref{090305_X+O_lc}.

\subsubsection{Optical observations}

\begin{figure*}
\includegraphics{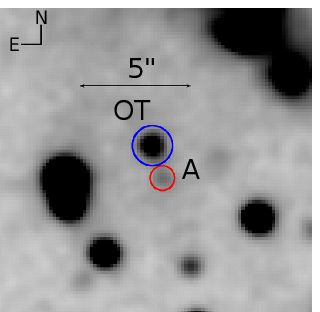}
\includegraphics{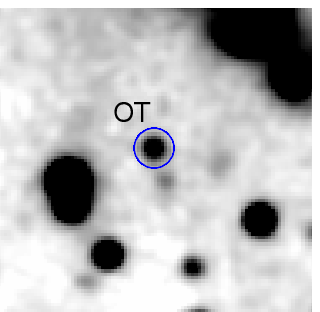}
\includegraphics{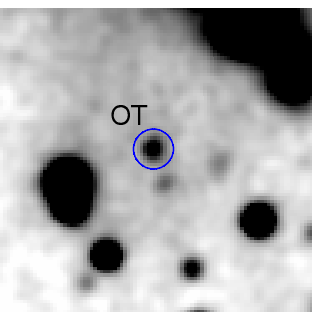}
\includegraphics{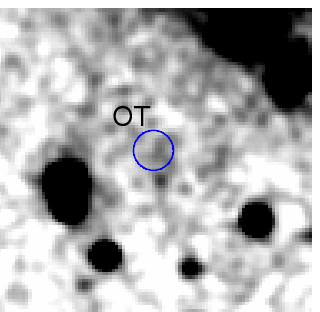}
\includegraphics{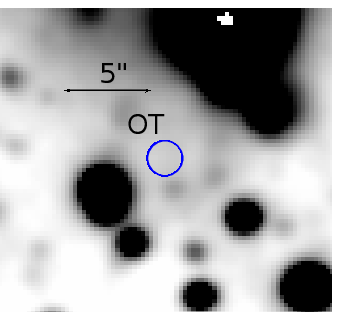}
\caption{
A finding chart for GRB\,090305A.
 All images except the final image are from Gemini $r^{\prime}$ band observations and show the optical transient (OT - blue circle) fading.
 No coincident host galaxy is found from late time Gemini imaging at the GRB position. The potential host galaxy, source A, is identified
 in the first epoch image with a red circle. Due to the faintness of source A, it is not possible to determine whether it is extended.
 The final image is from our additional late time observation using the FORS2 instrument on the VLT and also shows the lack of a coincident host galaxy
 down to the limits of the image.}
\label{090305_images}
\end{figure*}

Multiple observations of the field of GRB\,090305A were obtained using the Gemini South telescope
 as well as independantly detected with the GROND instrument on 2.2m telescope at La Silla Observatory (see \citealt{NicuesaGuelbenzu_2012_preprint_090305}
 where they also investigate the Gemini afterglow). 
 The optical afterglow was detected in all bands, as reported by \cite{090305_GMOS}, with observations being made 
 $\sim35$, $\sim55$ and $\sim75$ minutes after the GRB in the $r^{\prime}$, $g^{\prime}$, $i^{\prime}$ bands respectively. 
 Further observations were made in the $r^{\prime}$ band $\sim95$, $\sim125$ and $\sim13000$ minutes ($\sim9.02\,\mathrm{days}$)
 after the BAT trigger with the afterglow still detected in the first two epochs. The final Gemini epoch can be
 used to place a constraint on any host galaxy coincident with the GRB position, with the limit measured using an
 aperture equivalent to the full width at half maximum (FWHM) of the image. The $r^{\prime}$ band images are shown in figure \ref{090305_images}
 with the optical transient (OT) and nearby source A indicated. Variation of the afterglow flux density, $F$, is described using
 $F\propto t^{-\alpha}\nu^{-\beta}$. The spectral fit using the $r^{\prime}$, $g^{\prime}$, $i^{\prime}$ band detections has an index of $\beta_{O}\sim0.57$.
 The temporal fit to the $r^{\prime}$ band detections has slope of $\alpha_r = 0.44\pm0.02$, shown in figure \ref{090305_X+O_lc}, although this is a poor fit the
 limited number of points preclude the fitting of more complex models.


A final epoch of observations was made using the FOcal Reducer and low dispersion Spectrograph (FORS2)
 instrument on the Very Large Telescope (VLT) in the $R$ band. This was a long exposure image but due to the proximity of a group of bright stars
 could only marginally improve on the Gemini limit. Details of all observations made are listed in table \ref{090305_magnitudes}.

\begin{table*}
\begin{center}
\begin{tabular}{c c c c c c}
\hline
Start of observations & Exposure time & Mid-point $\Delta T\,(\mathrm{minutes})$ & Instrument & Filter & Magnitude \\
\hline
2009-03-05 05:47:23.5 & $5\times180\mathrm{s}$  & $36.87$     & Gemini-South GMOS & $r^{\prime}$ & $23.26\pm0.02$ \\
2009-03-05 06:07:00.6 & $5\times180\mathrm{s}$  & $56.51$     & Gemini-South GMOS & $g^{\prime}$ & $23.74\pm0.03$ \\
2009-03-05 06:26:41.5 & $5\times180\mathrm{s}$  & $76.19$     & Gemini-South GMOS & $i^{\prime}$ & $23.48\pm0.03$ \\
2009-03-05 06:46:21.2 & $5\times180\mathrm{s}$  & $95.84$     & Gemini-South GMOS & $r^{\prime}$ & $23.54\pm0.02$ \\
2009-03-05 07:07:20.5 & $4\times500\mathrm{s}$  & $125.53$    & Gemini-South GMOS & $r^{\prime}$ & $23.96\pm0.03$ \\
2009-03-16 05:37:04.7 & $10\times150\mathrm{s}$ & $\sim13000$ & Gemini-South GMOS & $r^{\prime}$ & $>25.69$       \\
2009-03-25 05:08:06.8 & $20\times240\mathrm{s}$ & $\sim28830$ & VLT FORS2         & $R$          & $>25.90$       \\ 
\hline
\end{tabular}
\caption{A log of Gemini and VLT observations of GRB\,090305A.
 Magnitudes quoted for the Gemini telescope are in the AB system and for the VLT are in the Vega system.
 These magnitudes have been calibrated from the standard Gemini zeropoints. Photometric errors are statistical only.
 All magnitudes have been corrected for Galactic extinction of $E(B-V)=0.22$ \citep{Schlegel_1998_Galactic_extinction}.}\label{090305_magnitudes}
\end{center}
\end{table*}

\subsection{GRB 091109B}
\subsubsection{Prompt and X-ray observations}

GRB\,091109B was detected by the BAT instrument on 2009 November 09 at 21:49:03 UT \citep{091109B_Swiftdetection}.
 The \emph{Suzaku} Wide-band All-sky Monitor (WAM), which also detected this GRB, measured an $E_{peak}=1330^{+1120}_{-610}\,\mathrm{keV}$
 showing the GRB is spectrally hard \citep{091109B_Ohno_Suzaku_WAM}. The GRB had a duration of $T_{90}=0.3\pm0.03\s$
 and a fluence ($15-150\,\mathrm{keV}$) of $1.9\pm0.2\times10^{-7}\,\mathrm{erg}\,\mathrm{cm}^{-2}$ \citep{091109B_T90}.
 As for GRB\,090305A, if we use a redshift of $z=0.3$ and extrapolate this fluence to the $1-1000\,\mathrm{keV}$ range
 we measure an isotropic equivalent energy of $E_{iso}=5.11\times10^{50}\mathrm{erg}$ 

\begin{figure}
\begin{center}
 \includegraphics{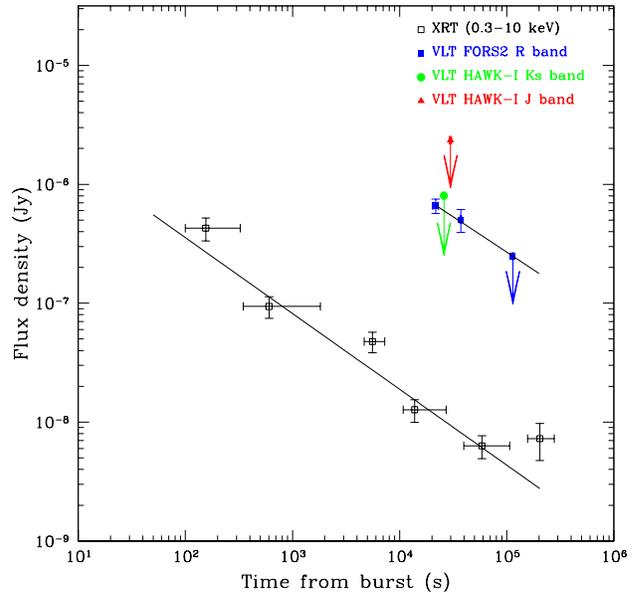}
\caption{
The XRT \citep{Swift_product1} and VLT optical light curves for GRB\,091109B. The X-ray data is plotted as black open squares and the $R$
 band optical measurements and limits are plotted as blue filled squares. Additional $J$ and $K$ band limits from the HAWK-I instrument
 on the VLT are plotted as a red triangle and a green circle respectively. Both the X-ray and the $R$ band data have been fit with power slopes
 giving decay indices of $\alpha_X=0.637^{+0.086}_{-0.084}$ and $\alpha_R=0.60\pm0.10$, consistent with each other.}
\label{X+O_lightcurves}
\end{center}
\end{figure}

The X-ray afterglow was detected by the XRT which began observing at 21:50:21.1 UT, $78.1\,\mathrm{s}$
 after the BAT trigger \citep{091109B_Swiftdetection}. 
 The X-ray light curve shown in figure \ref{X+O_lightcurves}, with data entirely taken in PC mode,
 can be fit by a power law with decay index $\alpha_X=0.637^{+0.086}_{-0.084}$ ($90\%$ errors).

\subsubsection{Optical observations}

We obtained multiple observations of the GRB position in the $R$ band using the FORS2 instrument and observations in the $J$ and $K$ band using
 the High Acuity Wide field K-band Imager (HAWK-I) instrument both on the VLT.

\begin{figure*}
\includegraphics{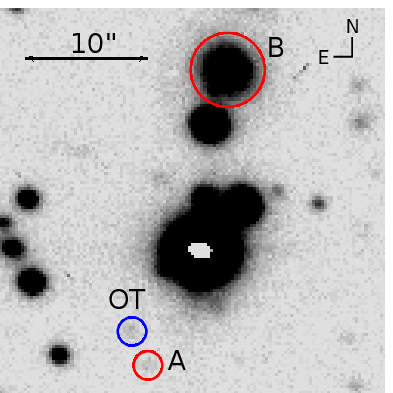}
\includegraphics{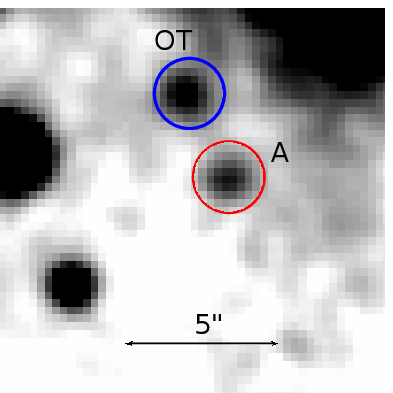}
\includegraphics{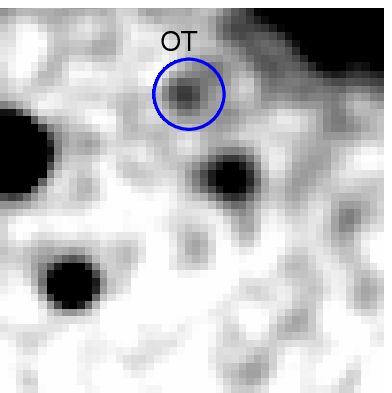}
\includegraphics{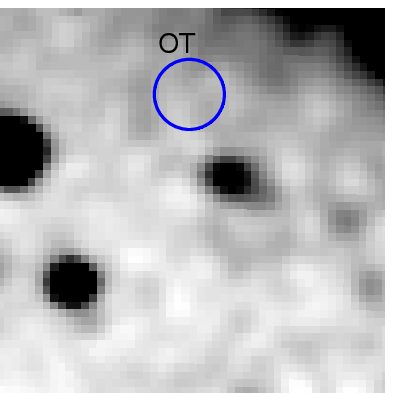}
\caption{
Finding chart for GRB\,091109B. The images were taken in the $R\_SPECIAL$ filter on the VLT. The first 2 panels are from the first epoch and indicate the optical transient (OT),
 a nearby source (A) and a bright galaxy (B). The optical transient can clearly be seen to be fading across the three epochs. Other objects visible in the field
 appear to be foreground stars and not galaxies.} \label{091109B_Rbandimages}
\end{figure*}

We discovered an optical afterglow in the $R$ band from two sets of $R$ band observations taken on 2009 November 11 $\sim360\,\mathrm{minutes}$
 and $\sim620\,\mathrm{minutes}$ after the BAT trigger with clear fading between the epochs
 \citep{Levan_GCN_091109B_VLT_obs,Malesani_GCN_091109B_OAG}. The  position was $\alpha=07^{\mathrm{h}}\,30^{\mathrm{m}}\,56.60^{\mathrm{s}}\pm0.02$,
 $\delta=-54^{\mathrm{o}}\,05\mathrm{'}\,23.3\mathrm{''}\pm0.3$, consistent with the revised X-ray position \citep{091109B_XRTposition}. 
 A third set of $R$ band observations were taken on 2009 November 11, $\sim1900\,\mathrm{minutes}$ ($\sim1.32\,\mathrm{days}$)
 after the start of the GRB and when the afterglow had faded allowing us to place a constraint on the magnitude of any
 underlying host galaxy. These images are shown in figure \ref{091109B_Rbandimages} with the optical transient (OT)
 indicated in the image along with two potential host galaxies: nearby faint source A and bright galaxy, source B.

$J$ and $K$ band observations were also made on 2009 November 10, $\sim495\,\mathrm{minutes}$ and $\sim430\,\mathrm{minutes}$
 after the GRB respectively. The transient was not detected in either band. The upper limit placed in the $K$ band
 implies that the afterglow emission was unusually blue, with a practically flat SED, and therefore suggestive of low extinction.
 We note that this also unusual in the context of the afterglow model where we would expect a steeper slope of at least $\beta=1/2$ \citep{Sari_fireball_AG}.
 All observations, magnitudes and limits are listed in table \ref{091109B_magnitudes} and shown in figure \ref{X+O_lightcurves}.
 The X-ray and optical temporal decay indices are $\alpha_X=0.59\pm0.05$ and $\alpha_R=0.60\pm0.10$, where the X-ray slope is consistent
 with that reported by \cite{091109B_XRTposition}.

\begin{center}
\begin{table*}
\begin{tabular}{c c c c c}
\hline
Start of observations   & Exposure time                 & Mid-point $\Delta T\,(\mathrm{minutes})$ & Band & Magnitude \\
\hline
2009-11-10 03:28:37.800 & $8\times300\mathrm{s}$        & 361.56  & $R$ & $24.13\pm0.14$ \\
2009-11-10 04:53:49.091 & $22\times10\times6\mathrm{s}$ & 431.95  & $K$ & $>22.23$       \\
2009-11-10 05:54:57.305 & $22\times10\times6\mathrm{s}$ & 497.45  & $J$ & $>21.99$       \\
2009-11-10 07:59:36.965 & $4\times300\mathrm{s}$        & 621.39  & $R$ & $24.42\pm0.21$ \\
2009-11-11 05:10:38.044 & $8\times300\mathrm{s}$        & 1903.28 & $R$ & $>25.63$       \\
\hline
\end{tabular}
\caption{A log of VLT FORS2 and HAWK-I observations of GRB\,091109B.
 Magnitudes are in the Vega system and have been corrected for Galactic absorption of $E(B-V) = 0.17$ \citep{Schlegel_1998_Galactic_extinction}.
 Photometric errors are statistical only.}\label{091109B_magnitudes}
\end{table*}
\end{center}

\subsection{GRB\,110112A}

\subsubsection{Prompt and X-ray observations}

\emph{Swift} detected GRB\,110112A with the BAT instrument on 2011 January 12 at 04:12:18 UT \citep{Stamatikos_GCN_110112A_Swiftdetection}.
 The duration of the GRB was $T_{90}=0.5\pm0.1\,\mathrm{s}$ and it had a fluence ($15-150\,\mathrm{keV}$) of
 $3.0\pm0.9\times10^{-8}\,\mathrm{erg}\,\mathrm{cm}^{-2}$ \citep{Barthelmy_GCN_110112A_BAT_refined}. 

An X-ray afterglow was detected by the XRT which started observing $75.5\,\mathrm{s}$ after the BAT trigger \citep{Stamatikos_GCN_110112A_Swiftdetection}. 

\subsubsection{Optical observations}

An optical transient was detected $\sim15.4\,\mathrm{hrs}$ after the BAT trigger in the $i^{\prime}$ band using the ACAM (Auxiliary-port camera)
 on the William Herschel Telescope (WHT). 
 This object was marginally coincident with the position for a candidate identified by \cite{Xin_GCN_110112A_OAG_detection}
 with the Xinglong TNT telescope, whose brighter magnitudes implies fading \citep{Levan_GCN_110112A_WHT}. This object was not detected in late-time imaging 
 using the GMOS instrument at the Gemini-South telescope confirming this object as the afterglow of the GRB. For further details of
 observations of GRB\,110112A, see Fong et al. (2012, in preparation). 

\begin{figure*}
 \includegraphics{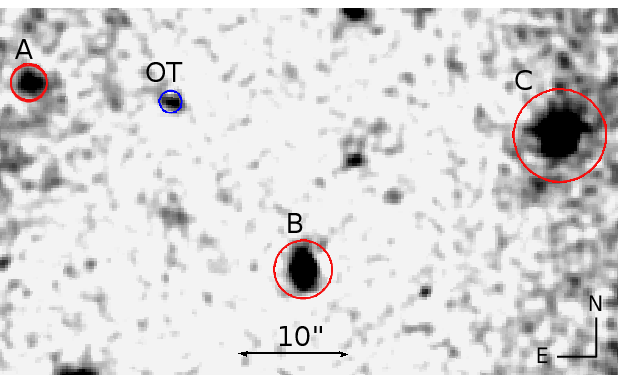}
 \includegraphics{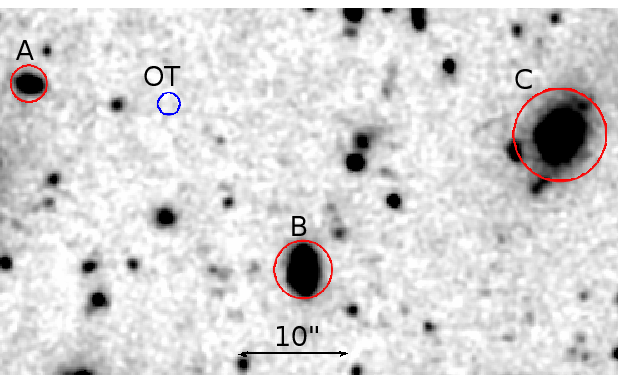}
\caption{Finding chart for 110112A. The images from the WHT ACAM instrument (left) and the Gemini-South GMOS instrument show the position of the optical transient (OT)
 and three potential host galaxies (A,B,C). These objects show evidence for extension.}\label{110112A_Findingchart}
\end{figure*}

\subsection{GRB\,111020A}

\subsubsection{Prompt and X-ray observations}

GRB\,111020A was detected by the BAT instrument on 2011 October 20 at 06:33:49 UT \citep{Sakamoto_GCN_111020A_Swiftdetection}.
 The duration of the GRB was $T_{90}=0.40\pm\,\mathrm{0.09}$ and it had a fluence ($15-150\,\mathrm{keV}$)
 of $6.5\pm1.0\times10^{-8}\,\mathrm{erg}\,\mathrm{cm}^{-2}$ \citep{Sakamoto_GCN_111020A_BAT_refined}.
 Extrapolating this fluence to the $1-1000\,\mathrm{keV}$ range, using a redshift of $z=0.3$, we measure
 an isotropic equivalent energy of $E_{iso}=6.15\times10^{49}\mathrm{erg}$. 

The X-ray afterglow was detected by the XRT which started observing $72.8\,\mathrm{s}$ after the BAT trigger \citep{Sakamoto_GCN_111020A_Swiftdetection}. 
 The afterglow was also observed by the \emph{Chandra} X-ray Observatory which placed the most
 precise position on the afterglow: $\alpha=19^{\mathrm{h}}\,08^{\mathrm{m}}\,12.49^{\mathrm{s}}\pm0.2$, $\delta=-38^{\mathrm{o}}\,00\mathrm{'}\,42.9\mathrm{''}\pm0.2$ 
\citep{Fong_2012_111020A_preprint_Xray}.

\subsubsection{Optical and Infrared observations}

Observing with the Gemini-South telescope in the $i^{\prime}$ band \cite{Fong_GCN_111020A_Gemini} noted
 the presence of several point sources near to the X-ray afterglow position. 
 However, with further imaging \cite{Fong_GCN_111020A_Geminiadd} later reported that none of these sources
 were fading between the epochs. Hence, no optical afterglow was observed for this GRB.

We obtained observations of GRB\,111020A with the VLT, equipped with HAWK-I in the $J$-band. Our observations
 started at 00:33 on October 21 2011, approximately 18 hours after the burst with a total exposure time of 44 minutes on sky. 
 We identify the same sources seen by \cite{Fong_2012_111020A_preprint_Xray} (in particular G1, G2 and G3). In our imagery we label
 these objects A (G3), B (G2), and C (G1) in order of increasing offset from the GRB position. We note that only object C 
 and perhaps B appears extended in our image with $0.4\mathrm{''}$ seeing.
 We do not identify any additional sources which are likely IR afterglow of GRB\,111020A to a limiting magnitude of $J=23.6$
 (3$\sigma$). 

We obtained late time observations of the GRB position in the $R$ band using the FORS2 instrument on the VLT.
 This observation is shown in figure \ref{111020A_Findingchart} and the details of the observation are given in table \ref{111020A_observations}.
 Although no optical counterpart was found, due to the small error (sub-arcsecond) on the \emph{Chandra} X-ray afterglow position
 we can still use this to accurately measure offsets from any potential host galaxies and to place deep limits at the position of the afterglow.

\begin{figure*}
 \includegraphics{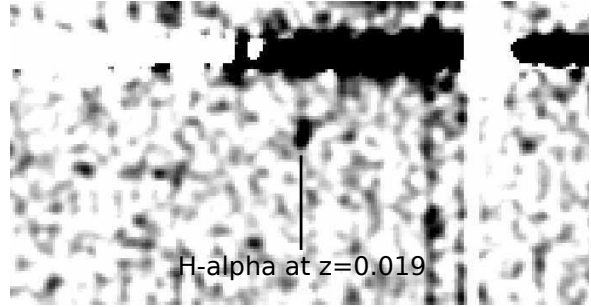}
 \caption{The VLT/FORS spectrum of the galaxy proximate to GRB 111020A.
 The source is significantly contaminated by an extremely bright foreground star sitting close to the location of the nucleus of the galaxy.
 We have attempted to remove this star by subtracting the median value across the wavelength range in each row from the spectrum.
 At an offset position, consistent with the disc of the galaxy we do observe an emission line at 6690\AA.
 If this line is interpreted as H$\alpha$, the inferred redshift is $z=0.019$}\label{GRB111020A_gal_spectrum}
\end{figure*}

\begin{figure*}
 \includegraphics{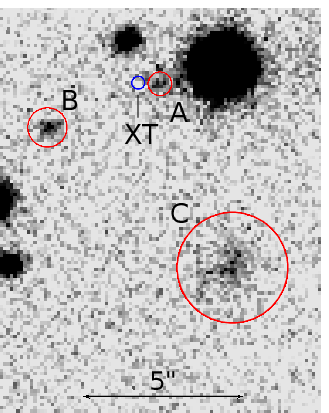}
 \includegraphics{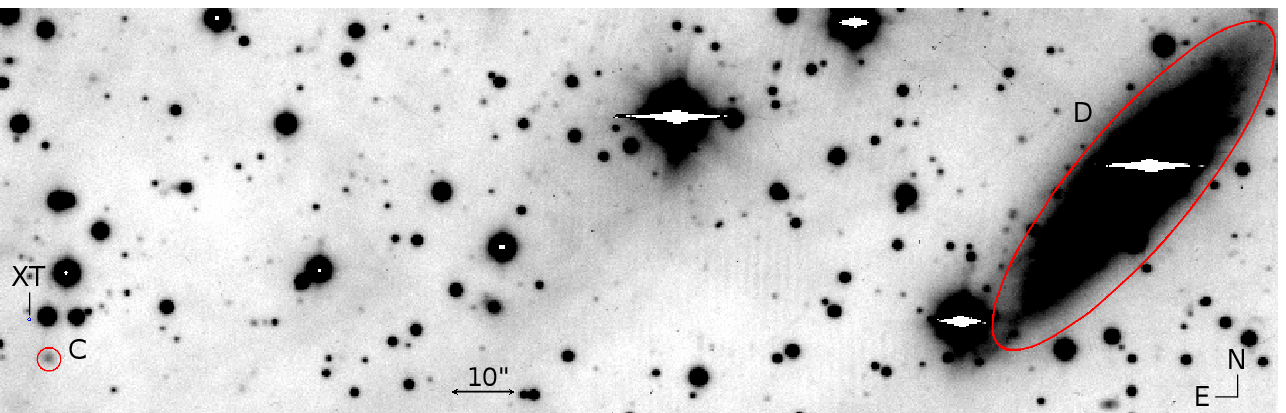}
\caption{VLT observations for GRB\,111020A. The early time $J$ band image is shown on the left with three nearby objects
 (A,B,C) shown in relation to the position of the X-ray transient (XT) with no infra-red transient detected. Objects A, B, C are synonymous with
 objects G3, G2, G1 respectively identified in \citet{Fong_2012_111020A_preprint_Xray}.
 Though objects B (G2) and C (G3) show evidence for extension, the faintness of object A (G1), the object closest to the
 X-ray position, makes this difficult to determine and the afterglow is still clearly offset from this object.
 In the FORS $R_{SPECIAL}$ band image shown on the right of the 3 objects we only detect galaxy C. In addition,
 we detect, a nearby large spiral galaxy (labelled D).}\label{111020A_Findingchart}
\end{figure*}

Additionally at this epoch we obtained FORS2 spectroscopy of a bright galaxy offset from the GRB position which is a plausible host 
 (see section 2.5) shown in figure \ref{GRB111020A_gal_spectrum}. The spectrum is significantly contaminated by light from a bright foreground star overlapping the galaxy and 
 confused with its nucleus. However, in the outlying regions of the galaxy we identify a weak emission line at 6688\AA. Identifying
 this as H$-\alpha$ suggests at redshift of 
 $z=0.019$, or $81\,\mathrm{Mpc}$.

\begin{center}
 \begin{table*}
  \begin{tabular}{c c c c c}
   \hline
   Start of observations  & Exposure time & Mid-point $\Delta T\,(\mathrm{days})$ & Band & Magnitude \\
   \hline
    2011-10-21 00:33:29.579 & $44\times10\times6\,\mathrm{s}$ & 0.77213 & $J$ & $>23.6$  \\
    2012-03-23 07:18:05.768 & $20\times240\,\mathrm{s}$       & 155.03  & $R$ & $>24.03$ \\
   \hline
  \end{tabular}
 \caption{A log of VLT HAWK-I and FORS2 observations of GRB\,111020A in the $J$ and $R\_SPECIAL$ filters. The HAWK-I observation was made at early times with no afterglow detected.
 Both observations allow us to place a constraint on any underlying host galaxy.
 The magnitudes quoted here are in the Vega system and have been corrected for Galactic absorption of $E(B-V) = 0.43$ \citep{Schlegel_1998_Galactic_extinction}.
 Photometric errors are statistical only.}\label{111020A_observations}
 \end{table*}
\end{center}

\subsection{Candidate host galaxies} 

For GRB\,090305A, GRB\,091109B, GRB 110112A and GRB\,111020A there is no host galaxy coincident with the optical afterglow position down to deep limits.
 Hence, we look investigate galaxies in the field around these SGRBs to determine if there are any strong host galaxy candidates.
 GRB\,090305A occurred in a region with a high density of sources meaning there are many field objects of the order
 of $5-8\mathrm{''}$ away, none of which show evidence of extension. There is a faint object located $1.48\mathrm{''}$ from the optical afterglow,
 shown in figure \ref{090305_images} and labelled as source A. The magnitude of source A is $r^{\prime}=25.64\pm0.20$. 
 Due to its faint magnitude it is not possible to determine with high confidence whether this object is extended.

For GRB\,091109B we identified two extended objects which are potential host galaxies.
 The first is a faint object $\sim3.0\mathrm{''}$ from the GRB position marked in figure \ref{091109B_Rbandimages} as 
 source A with $R=23.82\pm0.10$. The object is not detected in the $J$ and $K$ band down
 to the limiting magnitude listed in table \ref{091109B_magnitudes}.
 The second potential host, source B, is a spiral galaxy located $\sim22.5\mathrm{''}$ from the GRB position.
 Even though this is located much further from the GRB position  than source A, it is much brighter with magnitudes
 $R=19.19\pm0.05$, $J=17.45\pm0.05$ and $K=16.16\pm0.05$.

GRB\,110112A occured in a relatively clear region quite well offset from other objects in the field.
 We find 3 spiral galaxies which could potentially be host galaxies labelled as A, B and C in figure \ref{110112A_Findingchart}.
 These objects have $i^{\prime}$ band magnitudes of $22.70\pm0.07$, $21.16\pm0.04$ and $20.17\pm0.05$ respectively with
 increasing offsets of $12.9\mathrm{''}$, $19.3\mathrm{''}$, $35.6\mathrm{''}$. 

In our images we identify four objects which could be candidate host galaxies for GRB\,111020A, shown in figure \ref{111020A_Findingchart}.
 Three of these objects are within $7\mathrm{''}$ of the X-ray afterglow with the fourth being a large, extremely bright spiral galaxy at an offset of $166\mathrm{''}$.
 The labels of A-D are with increasing offset from the afterglow position.
 This galaxy, labelled D in our image, has a magnitude $R\sim14.0$ with the uncertainity due to a number of saturated foreground stars obscuring
 the galaxy, making it difficult to make a precise measurement. Our measured redshift for the galaxy (if taken from the single line)
 is $z=0.018$, corresponding to a projected separation of $60\,\mathrm{kpc}$.
 The 3 objects nearby have offsets of $0.7\mathrm{''}$, $3.0\mathrm{''}$ and $6.8\mathrm{''}$ respectively, labelled as A, B, C in figure \ref{111020A_Findingchart}
 corresponding to galaxies G3, G2 and G1 in \cite{Fong_2012_111020A_preprint_Xray}.
 All these objects are detected in our HAWK-I data with magnitudes $J=22.00\pm0.09$, $21.41\pm0.10$ and $20.70\pm0.08$ respectively,
 but with only objects B, C showing evidence for extension. However, only object C is clearly detected in our FORS data with $R=21.34$ and object B marginally detected.

To give an indication of whether these galaxies are strong candidates to be the host we ask what is the probability, $P_{\rm chance}$,
 that an unrelated galaxy of the same magnitude or brighter would be found within the given offset.
 This approach has been considered extensively for LGRBs and SGRBs \citep[e.g.,][]{GRBoffset,Levan_2008_050906_BAT_brighthost,Berger_2010_nohost}.
 Here we use a simplified version based on the offset between the SGRB position and the given galaxy and compare this to the distribution of such offsets in a large sample
 of random sky positions, as described in section~\ref{diagnostic_tool_section}. 
 For source A associated with GRB\,090305A we measure a value of $P_{\rm chance}=0.09$ which, 
 though a low value, does not provide a firm host association. Similarly, for GRB\,091109B sources A and B have values of
 $P_{\rm chance}=0.09$ and $0.10$ respectively.
 For GRB\,110112A all three host candidates have relatively higher $P_{\rm chance}$ values of $0.50$, $0.34$ and $0.42$ with galaxy B having the lowest $P_{\rm chance}$
 value.

 Looking at the objects we detect in the field of GRB\,111020A, in both our $R$ and $J$ band data, we find $P_{\rm chance}$ values of $0.007$, 
 $0.04$, $0.09$ and $0.05$ for objects A (G3), B (G2), C (G1) and D. If object A (G3) is indeed a galaxy then this object
 is strongly associated and hence the GRB could be classified as ``hosted''. However, we caution that this field is at low galactic latitude
 and that this object could be faint star. In addition, this object does not show significant evidence for extension and so we do not conclusively rule out the other galaxies
 as potential hosts.

\section{Identifying hostless SGRBs}

\subsection{A sample of SGRBs}\label{SGRB_sample}

We utilise a sample of all SGRBs detected up to April 2012. This includes bursts with $T_{90} < 2\,\mathrm{s}$, and those bursts which have been declared
 short bursts with extended emission by the {\em Swift} team in the GCN circulars archive. Since we are interested in host identifications
 we further cull this input list to require at least an XRT detection of the afterglow, since BAT-only positions are insufficient
 to identify hosts with moderate to high confidence unless they are extremely bright (e.g. \citealt{Levan_2008_050906_BAT_brighthost}).
 In addition, since we are interested in the host galaxies of these SGRBs, we do not list XRT-localised SGRBs where a host galaxy search has not been reasonably attempted.
 By these criteria the sample includes 40 GRBs: 33 SGRBs with $T_{90} < 2\,\mathrm{s}$ and 7 with extended emission.
 Of these GRBs 26 are well-localised and 14 are localised using the XRT.
 Our complete host galaxy sample along with some basic properties of the GRBs is shown in Table~\ref{OAG_hosts} and ~\ref{XAG_hosts}. 

\subsection{Difficulties of host identification}

Probabilistic arguments of the sort outlined above have been used to
 argue for a host galaxy associations for many {\em Swift} SGRBs
 (e.g. \citealt{Prochaska_2006_SGRBhosts,Kocevski_2010_070724A_analysis,Rowlinson_2010_080905A_Opt,McBreen_2010_090510_LAT_analysis}).
 However, $P_{\rm chance}$ is calculated as an independent quantity for each galaxy considered, meaning that for some SGRBs
 there are several galaxies with roughly similar $P_{\rm chance}$ values. For example, there may 
 frequently be multiple galaxies in the field for which $P_{\rm chance} < 5\%$ \citep[e.g.][]{AL_random_associations}, but in these cases this clearly does not
 represent the true false positive rate for a given burst (since at most one can be the true host). This potentially produces a degeneracy in identification
 of candidate hosts between bright galaxies with large offsets and faint (or even undetected) galaxies with small offsets \citep{Berger_2010_nohost},
 as seen for GRBs\,091109B and 111020A.

\begin{figure*}
 \begin{center}
  \includegraphics{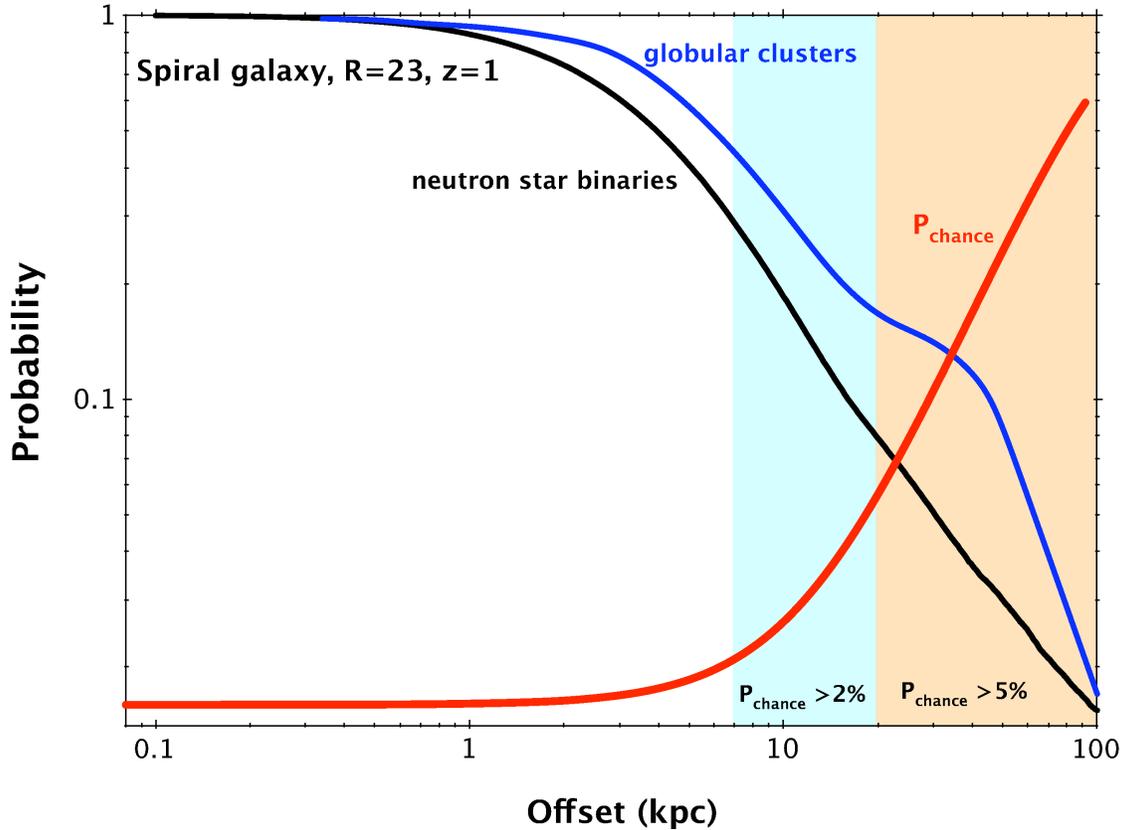}
 \end{center}
 \caption{This plot shows the distribution of NS binary mergers with respect to a Milky Way-like spiral galaxy at $z=1$.
 This is shown for both the primoridal and the dynamical channel of NS binary formation. The value of $P_{\rm chance}$ we would expect
 for this galaxy at $R=23$ are also shown with the 2\% and 5\% $P_{\rm chance}$ areas highlighted.} \label{NS_NS_P_Chance} 
\end{figure*}

Such problems may be largely unavoidable for any population of events that arises at large offsets, however our prior expectations may
 impact the likelihood assigned to either the kicked or high-redshift (and undetected host) scenarios. For example, the association
 of LGRBs with massive star collapses mean that the non-detections 
 of hosts to deep limits in LGRBs are typically ascribed to faint host galaxies, with the expectation that
 deep observations will ultimately find them underneath the burst positions \citep[e.g.][]{Hjorth_2012_TOUGH_catalog,Tanvir_2012_HST_LGRB_hostgalaxies}. For short GRBs the models
 are far less well constrained. It is clear that within the population that lie on host galaxies, the positions are not associated
 strongly with active star formation \citep{Fong_2010_SGRBhosts_HST}, and therefore the progenitors can likely have longer lifetimes than those of 
 LGRBs. The population of SGRBs identified to date also has a much lower mean redshift than for LGRBs ($<z>\sim0.7$ compared to $<z>\sim2.2$)
 \citep{Jakobsson_2006_mean_redshift_LGRB,Jakobsson_2012_TOUGH,Fong_2010_SGRBhosts_HST},
 if this distribution is representative of the underlying distribution then it is more likely that observations of a given depth would
 uncover the host galaxy (see section \ref{redshift_gal}). Given
 this it is natural to also consider the difficulties that may arise identifying hosts from a kicked population.

To demonstrate the difficulties
 in the kicked scenarios, we can consider the likely implications of a binary merger model for the locations around a given galaxy.
 Indeed, if NS binary mergers are indeed responsible for the creation of SGRBs then we may expect larger offsets due to both the natal kick
 and the delay in merger time after creation (potentially $>10^{10}\,\mathrm{yrs}$, although some authors have argued for much shorter delays for
 the bulk of the population \citep{Belczynski_2006_NSmerger}.
 Figure \ref{NS_NS_P_Chance} shows typical NS binary merger sites with respect to a spiral host galaxy with $R=23$ at $z=1$, comparable in 
 luminositiy ($M_{B}\sim -20.4$) to the Milky Way (from \citealt{Church_NS_offsets}). 
 We include both the case of NS binary systems formed through a primordial channel and a dynamical channel within globular clusters (GCs),
 where a neutron star captures a companion through three-body interactions \citep{Grindlay_2006_SGRBs_GCs} (i.e we also plot
 the distribution of globular clusters).
 We compare this with the $P_{\rm chance}$ values we would measure as a function of projected offset (in kpc) for this particular case of a galaxy with $R=23$ at $z=1$.
 We also highlight the areas where $P_{\rm chance}$ is 2\% and 5\%.
 This demonstrates that for this model of NS binary mergers we may expect a significant fraction to merge at a point where we can no longer associate the
 resulting explosion with the host galaxy from which it originated (see also \citealt{Berger_2010_nohost}). This problem becomes even more
 acute when considering fainter galaxies. A lower luminosity galaxy at the same redshift will have significantly higher 
 $P_{chance}$ at moderate offsets, making a firm association difficult. 
 In other words, in any kicked scenario we would expect a significant
 fraction of bursts for which we cannot identify any host galaxy. 
 The implications this could have for hostless GRBs are further discussed in section \ref{kicks}.
 

\subsection{A diagnostic tool to help investigate short burst host associations} \label{diagnostic_tool_section}

 
 Given the above limitations we take an alternative approach to calculating $P_{\rm chance}$, by directly calculating for multiple random positions on the sky.
 We took
 15,000 random positions within the footprints of the Sloan Digital Sky Survey (SDSS) \citep{Abazajian_2009_SDSS_Seventh_Data_Release}
 and the Cosmological Evolution Survey (COSMOS) \citep{Capak_2007_COSMOS_data_release},
 and used the measured galaxy data available within these surveys to find the nearest galaxies brighter than a given magnitude.
 The galaxy magnitude range chosen was reflective of SGRB host galaxies, using SDSS data at the bright end ($15-22\,\mathrm{mag}$)
 and supplementing this with data from the Subaru telescope in the COSMOS survey at the faint end ($>22\,\mathrm{mag}$).
 We excluded any SDSS positions with Galactic extinction values $E(B-V)>0.1$ and corrected for Galactic extinction for all galaxy magnitudes within the sample. For the COSMOS sample we utilize galaxy magnitudes measured within a $3^{\prime \prime}$ fixed aperture since for
 faint galaxies this should enclose essentially all of the light. 

\begin{figure*}
 \begin{center}
   \includegraphics{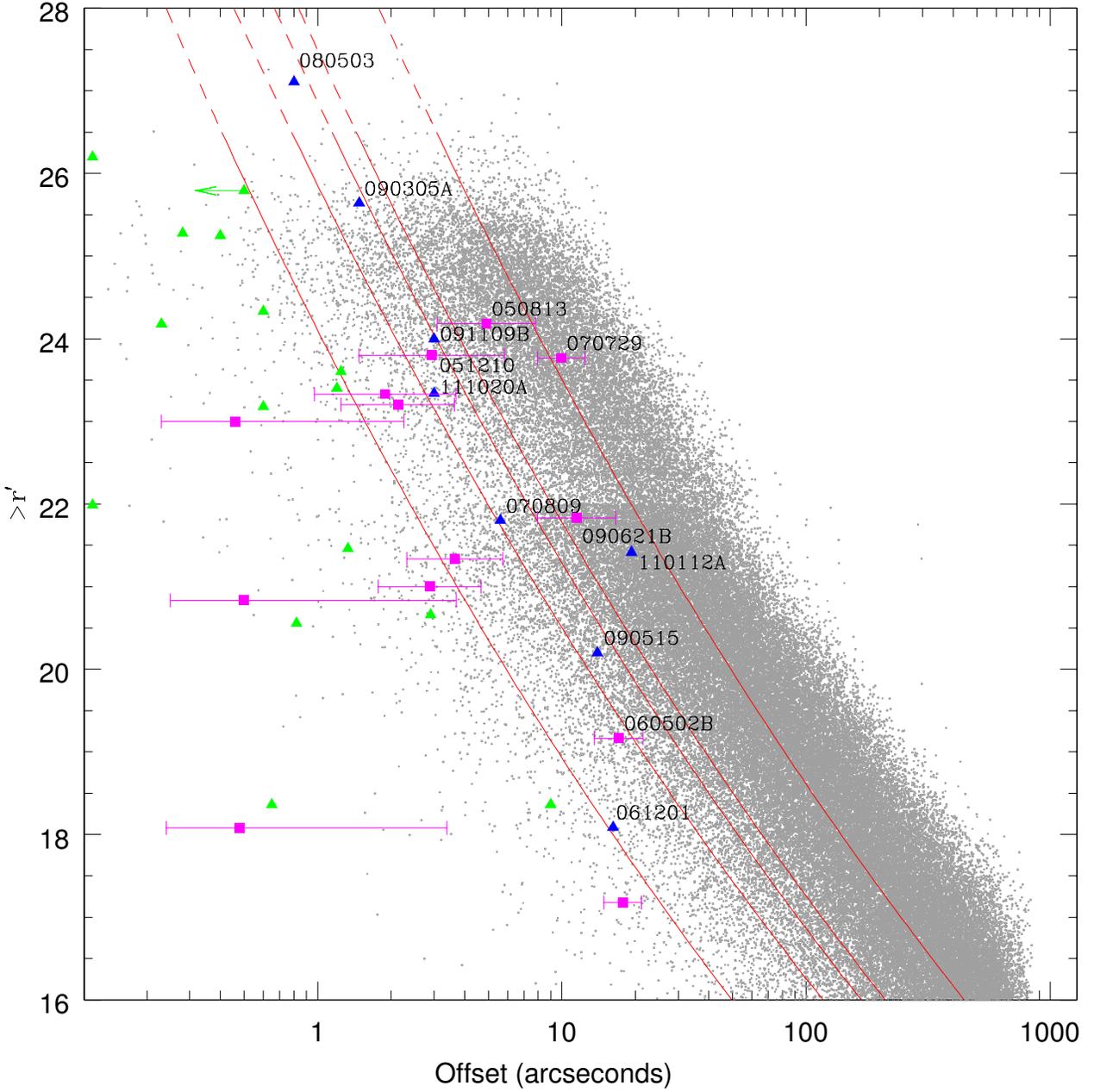}
 \end{center}
\caption{Using random positions in the Sloan Digital Sky Survey (SDSS) and the Subaru telescope in the Cosmological Evolution Survey (COSMOS)
 we have plotted the minimum offsets for galaxies brighter than a given $r^{\prime}$ magnitude for 15,000 random positions.
 The solid red lines are the 1\%, 5\%, 10\%, 15\% and 50\% percentiles for this dataset, with the dotted lines showing the extrapolation of the fit beyond our dataset. 
 We have plotted the galaxies with the lowest $P_{\rm chance}$ for all SGRBs with optical (or CXO) localisations (triangles)
 and for SGRBs with only {\em Swift}/XRT positions (purple squares). The blue triangles represent galaxies in the field of hostless SGRBs,
 with only one galaxy being included for each GRB. All well-constrained host galaxies (green triangles)
 lie below the 1\% percentile line.} \label{Offset_gal} 

\end{figure*}

The galaxy distribution is shown in figure \ref{Offset_gal} along with the enclosing percentiles (1\%, 5\%, 10\%, 15\% and 50\% percentiles)
 of the distribution, i.e. beyond the 1\% line only 1\% of the galaxies in the sample have magnitude $<r^{\prime}$ and are at offset $<$ observed.
 Also shown are the putative host galaxies of the SGRB population. We would expect any well constrained host galaxy (by $P_{\rm chance}$ or
 similar probabilistic measure) to be an outlier to the main galaxy distribution and, indeed, we see that all strongly associated SGRB
 host galaxies lie below the 1\% percentile line. 
 For hostless GRBs, however, the proposed host galaxies lie closer to the background galaxy distribution.

We can use the division between GRBs strongly associated with their host galaxies and those considered hostless shown on figure \ref{Offset_gal} to define
 an `association radius', $\delta x$. For a galaxy of a given magnitude, this is the offset from the centre of the putative host galaxy
 within which we can say the GRB is strongly associated, specifically there is a less than 1\% chance that an unrelated galaxy of this magnitude
 (or brighter) would appear within this distance. 
 Conversely, if a GRB does not fall within this radius for \emph{any} nearby galaxy then we chose to describe it as hostless.
 $\delta x$ (in arcseconds) is given by equation \ref{1_percentile_fit}.

\begin{center}
\begin{align}
\delta x = 1.48\times10^{13}\,m_{r}^{-\gamma}
\label{1_percentile_fit}
\end{align} 
\end{center}

where $m_{r}$ is the $r^{\prime}$ band magnitude of the suggested host galaxy and 
 $\gamma=9.53$
 describes the best-fit power law dependence and is an empirical fit to the data. Hence, we can define a hostless short GRB
 where the position of the GRB (allowing for the error box) does not fall within the association 
 radius of any nearby galaxies. The choice of association radius clearly depends on
 the minimum probability that one is prepared to accept (e.g. one could prescribe differing 
 radii at different confidence levels). However, the 1\% contour is broadly 
 applicable, and would suggest for our sample of $\sim 20$ SGRBs we may expect one
 to be falsely associated with a galaxy which is not its host. Using this approach
 we identify a total of seven (eight) hostless SGRBs, 061201, 070809, 080503, 090305A, 090515, 091109B, 110112A (and potentially 111020A).
 We also note that GRB\,111117A is very close to the 1\% boundary \citep{Margutti_2012_111117A_Chandra_host}.

We also note that 4 of the XRT-localised GRBs are hostless (i.e. the error bars due to the
few arcsecond XRT positions do not cross the 1\% line): 050813, 051210, 060502B, 070729 and 090621B.
 For GRB\,090621B, however, this may be due to the lack of a deep search, thus far.

\begin{figure*}
 \begin{center}
  \includegraphics{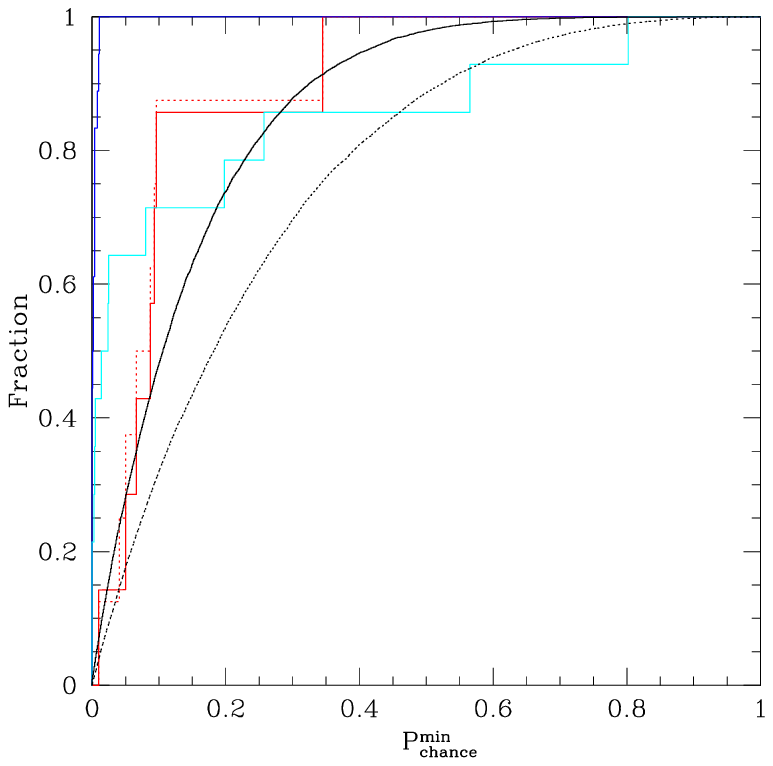}
  \includegraphics{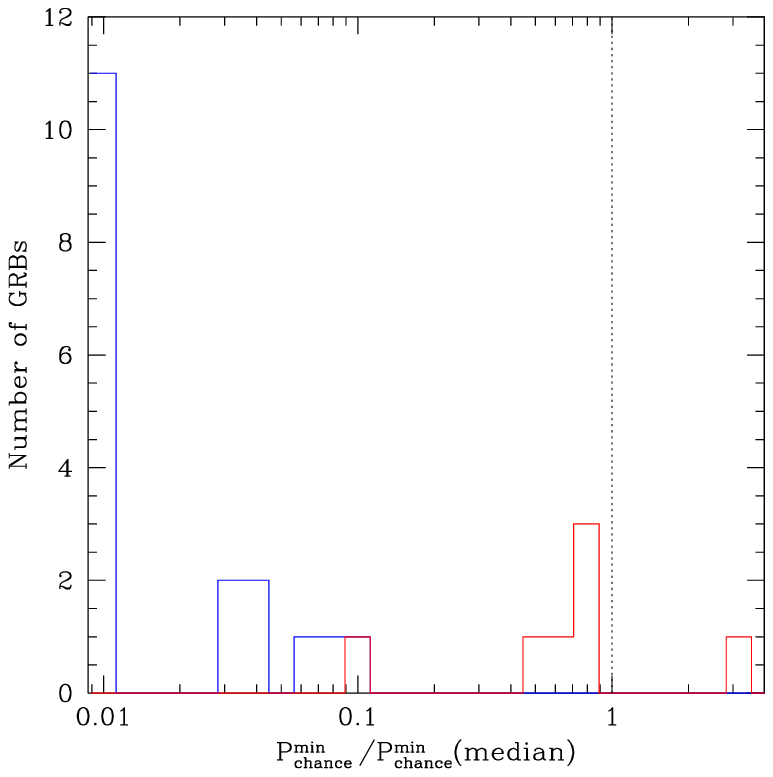}
 \end{center}
 \caption{The plot on the left shows the cumulative distribution of $P_{\rm chance,min}$ for a set of 15,000 random positions within the SDSS and COSMOS footprints (black line).
 For comparison, the $P_{\rm chance,min}$ values calculated using the method of \citet{GRBoffset} (dotted black line) are included. This is clearly distinct $P_{\rm chance}$
 values calculated with the method outlined in the text.
 Also shown is the cumulative distribution of $P_{\rm chance,min}$ for the 
 well-localised SGRBs with a host galaxy (blue), hostless SGRBs (red) and the {\em Swift}/XRT only detected SGRBs (cyan).
 The dotted red line is the hostless sample with the inclusion of a our identified galaxy (galaxy B) for GRB\,111020A, but this is uncertain due to the presence of
 another object close-by which could be a faint star as explained in the text.
 The plot on the right shows a histogram of the ratios of the $P_{\rm chance,min}$ values to the median of the random galaxy distribution,
 for all the well-localised SGRBs in our sample. As for the plot on the right the hostless GRBs are shown in red and the other well-localised GRBs in blue.
 A KS test of these distributions compared to the random positions does not show that these populations are statistically distinct.
 However, looking at the median values does show that a significant fraction (6/7) of secure hostless cases the $P_{chance,min}$ values are less than the median value
 for the random galaxy distribution.
 } \label{Rand_gals_P_chance_min}
\end{figure*}

We can now compare the galaxies with the lowest $P_{\rm chance}$ values ($P_{\rm chance,min}$) for our seven (eight with our potential host
 of GRB\,111020A) bursts to the distribution of random locations on the sky.
 This will allow us to ascertain if this sample of events, originate from
 the underlying galaxy distribution, as might be expected for local, but kicked SGRBs, 
 or are essentially uncorrelated, as might be expected for SGRBs originating from
 higher redshift.
 To do this for each random location we identify the galaxy
 with the minimum $P_{\rm chance}$, and hence arrive at a distribution of $P_{\rm chance,min}$ for
 all random locations.
 Measuring the $P_{\rm chance}$ value is done by calculating the lines for multiple percentiles using the same formulaic form as equation \ref{1_percentile_fit}
 but varying the coefficient values. The percentile lines themselves can then be used to calculate the $P_{\rm chance}$ values.
 This is shown graphically in Figure \ref{Rand_gals_P_chance_min}\footnote{
 Note that although $P_{\rm chance,min}$ is strictly a function of the galaxy magnitude range considered,
 the span of galaxies covered by our joint SDSS/COSMOS analysis is fairly representative of
 the typical range accessible in our GRB host fields}.
 The KS probability that the hostless SGRB positions are drawn from an underlying
 random distribution of the sky is $P_{KS} = 0.19$, which is not a rejection of this statement. We also show in Figure~\ref{Rand_gals_P_chance_min}
 the curve that would be obtained from a calculation of $P_{\rm chance,min}$ via traditional routes utilising only the number counts. 
 The resulting curves are clearly different, this is particularly true at the low $P_{\rm chance,min}$ range, where the two curves differ by almost
 a factor of two. Formally, the probability of this latter curve matching the observations is only $P_{KS} = 0.01$, which is a rejection of the statement
 at $\sim 3\sigma$.  In other words, the use of number counts alone can lead to an underestimate of the real probability of chance alignment of a galaxy with
 a given position on the sky. 
 
As an alternative, we compare the $P_{\rm chance,min}$ values for the hostless GRBs to the median of the random galaxy distribution, looking only at galaxies
 above the limiting magnitude placed on each individual hostless GRB, we find that in 6/7 secure hostless cases (i.e. excluding GRB\,111020A) the $P_{\rm chance,min}$ value
 is less than the median.
 Only in the case of GRB\,110112A do we find that the $P_{\rm chance,min}$ value is greater than the median.  For a random sky distribution we would expect to sample
 this distribution evenly and the binomial probability of 6/7 (7/8) events lying at less than the median is 0.0547 (0.0313),
 again perhaps indicating that we are observing SGRBs from local structure in their fields.
 
 Hence, these results are suggestive that we are observing SGRBs that are correlated with large scale structure on the sky, 
 even if the individual host galaxy that we identify with $P_{chance,min}$ is not the true host. 
 This may reflect that these bursts are kicked from relatively low redshift, and relatively luminous 
 galaxies. However, it is also possible that we are observing these SGRBs from
 moderately massive structures at higher redshift, and hence the low values
 for $P_{chance,min}$ are actually reflecting other cluster members. Early estimates from \cite{Berger_2007_SGRBs_clusters} put the
 percentage of SGRBs in clusters at $\sim20\%$ (e.g. \citealt{Bloom_2006_050509B,Levan_2008_050906_BAT_brighthost}), though further studies of this nature have yet to be published.

 
Furthermore, the galaxies identified with lowest $P_{\rm chance}$ in these cases, are likely to be luminous (due to Malmquist bias), and may be massive
(for an assumed mass to light ratio for a spiral galaxy),
such galaxies are more likely to retain any dynamically kicked systems
 within their haloes. In contrast, lower mass galaxies have a much larger escape fraction, and
 are more likely to create a population of hostless SGRBs (see also \citealt{GRBoffset,FryerNS_dist}). Alternatively, the relative offsets
 from the galaxies with lowest $P_{\rm chance}$ means at least some GRBs may be residing within  
 globular cluster systems within these hosts.
 In this case it may be that we are observing
 a population of dynamically formed binaries within these globular clusters, and that, given
 the typically large distances we cannot directly observe the host cluster 
 (e.g. \citealt{Grindlay_2006_SGRBs_GCs,Salvaterra_Swift_Bimodal,Church_NS_offsets}).
  \cite{Grindlay_2006_SGRBs_GCs} predicts $10-30\%$ of SGRBs could be explained in dynamical mergers, however, as shown in 
  Figure~\ref{NS_NS_P_Chance} the majority of globular clusters are not at such large radii. Hence, while some of the hostless
  systems could arise from this channel it seems unlikely that all of them would. 

\section{Discussion}\label{Disc_section}

By considering the locations of SGRBs relative to random locations on the sky we have shown that even those SGRBs for which
 it is most difficult to unambiguously identify a host are likely to be at moderate redshift. 
 However, it is possible that rather than observing the hosts themselves, we instead are tracing
 larger scale structure.

\subsection{Constraints on the possibility of high-redshift host galaxies} \label{redshift_gal}

We can use the $3\sigma$ point source detection limits at the locations of the hostless GRBs in our sample to place constraints on the 
 possibility of a coincident host galaxy. 
 For increasing redshift 
 we can determine the minimum galaxy luminosity, $L_{lim}$ as a fraction of $L^*$ \citep[the characteristic luminosity of the knee of the
 Schecter luminosity function; ][]{Schechter_L_fn}, we would be able to detect in a given filter. 
 Using this $L_{lim}$ we can determine the probability of detection of a galaxy, $P_{detect}$, under the simplifying assumption that the 
 likelihood of a galaxy producing an SGRB is proportional to its luminosity.\footnote{But see also \cite{Belczynski_2006_NSmerger} } Specifically, $P_{detect}$ describes the fraction of
 the luminosity-weighted luminosity function that we probe with increasing redshift.
 To perform this analysis we used the templates for Sbc, Elliptical (Ell) and Irregular (Irr) galaxies from \cite{Galaxy_templates} and considered magnitudes in the
 SDSS $r^{\prime}$ band. We note that the template used is based on an agregate of observed local galaxies and so includes typical levels of extinction 
 appropriate for the galaxy type. In addition, we have neglected the effect of surface brightness dimming, but we expect this to be 
 a small effect for these sources which are only marginally resolved. 
 For more detailed information on this analysis we refer the reader to Appendix \ref{Appendix_P_detect}.

\begin{figure*}
\includegraphics{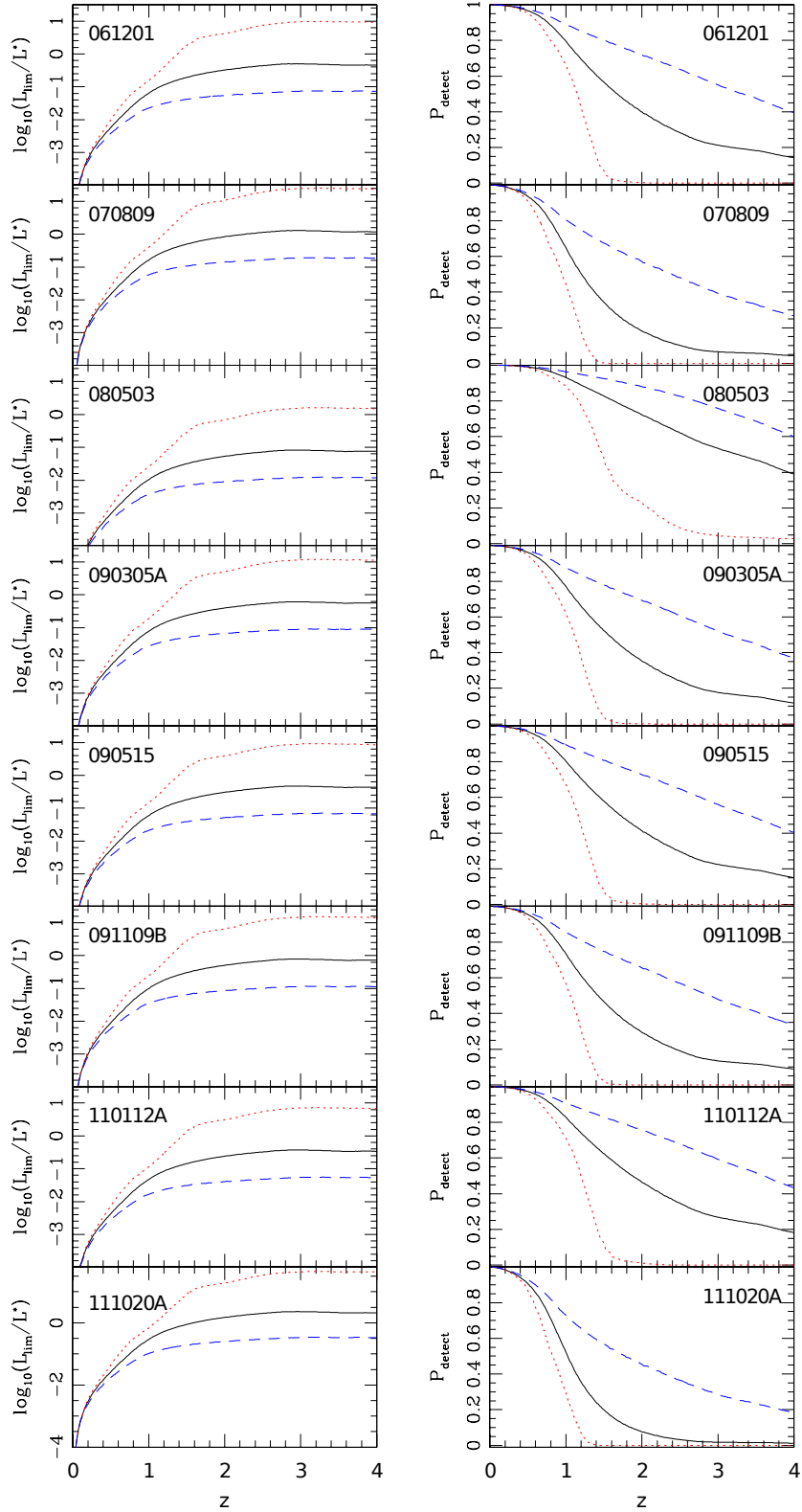}
\caption{The set of panels on the left hand side show the minimum luminosity galaxy, $L_{lim}$, we could detect as a function of redshift in the SDSS $r^{\prime}$
 band and based on the magnitude limits of all the hostless GRBs in our sample. Each of these panels contains this evolution for an Sbc (black, solid line), Ell (red,
 dotted line) and Irr (blue, dashed line) galaxy. We use our measured deepest $r^{\prime}$ limits placed on GRB\,090305A, 091109B
 (both adapted from the VLT $R$ band limits) and 110112A (extrapolated from the Gemini-N $i^{\prime}$ band).
 We have also included published limits for GRBs 061201, 070809, 080503, 090515 and 111020A \citep{Perley_2009_080503_gals_submitted,Berger_2010_nohost,Fong_2012_111020A_preprint_Xray}
 again adapted to the $r^{\prime}$ band where necessary. There is a deeper limiting magnitude for GRB\,111020A from \citet{Fong_GCN_111020A_Gemini}
 and hence this value is used here rather than our reported limit.
 Note that this includes a simple linear evolution of the luminosity function with further details in Appendix \ref{Appendix_P_detect}.
 Using these $L_{lim}$ values, we then plot the fraction of the integrated luminosity we probe at increasing redshift in the right hand set of panels ($P_{detect}$).
 This is again shown for the 3 types of galaxies we consider here.
 }\label{L_lim_P_detect}
\end{figure*}

Figure \ref{L_lim_P_detect} shows this analysis using the detection limits determined for all the hostless GRBs in our sample.
 Looking at the redshift range thought to be typical for SGRBs ($z<1$)
 we find for at least six of the bursts we uncover $\sim85\%$ of the integrated galaxy luminosity for the
 Irr galaxy type, $\sim73\%$ Sbc galaxy type and $55\%$ for the Ell galaxy type. In addition, the extremely deep limit placed on GRB\,080503 means that
 even for the Ell galaxy type we uncover $88\%$ of the integrated galaxy luminosity. 
 This analysis suggests that it is unlikely that the hostless SGRBs are simply cases of faint, but coincident,
 hosts in the redshift range for $z<1$, though this case is less strong for Ell galaxies. Higher redshifts, however, up to e.g., $z\sim4$,
 are not ruled out by observations of the afterglows or the limits on coincident host galaxies.

Though the majority of SGRBs have been found at redshifts in the range $0.1-0.9$ there is some evidence indicating the existence of
 a higher redshift population. GRB\,090426 has the highest confirmed redshift ($z=2.609$) of a GRB with $T_{90}<2\s$
 \citep{Levesque_2010_090426_host,Thone_2011_090426_environ}.
 However, there remains considerable discussion as to its nature with its host galaxy, environment and spectral properties being more suggestive
 of a massive star progenitor than a compact binary merger despite its short duration
 \citep{Antonelli_2009_090426, Levesque_2010_090426_host, Xin_2010_090426_Opt, Thone_2011_090426_environ}.
 There is potential evidence for other SGRBs at high redshift with indications for a $z>4.0$ host for GRB\,060121
 \citep{Levan_2006_060121_AG+host, deUgartePostigo_2006_060121_highzhost}, beyond the redshift detection limits we have placed for a fainter galaxy,
 and the faint, possibly high redshift hosts of GRB\,060313 ($z\lesssim1.7$) \citep{060313_zlim_GCN} and GRB\,051227 \citep{Berger_2007_highzSGRBhosts}.
 In particular, GRB\,070707 ($z<3.6$), has a very faint coincident host at $R\sim27$ \citep{Piranomonte_2008_070707_Beta-OX}. 
 These results may indicate that 
 hostless GRBs are a window onto a higher redshift population. However, some hostless SGRBs have been probed to deep limits,
 especially GRB\,080503 with a limit of $\mathrm{F606W}>28.5$ from \emph{HST} \citep{Perley_2009_080503_gals_submitted}.
 If the delay-time distribution for SGRB progenitors is long, then we would expect to preferentially see them at lower redshift.
 Under the binary neutron star model a significant component will be created around the peak of the Universal star formation rate
 and so will merge in a time $\sim10^{9}\,\mathrm{yrs}$ after this era \citep{OShaughnessy_2008_merger}. 

A further constraint on the high redshift scenario can be obtained simply by contrasting the properties of the long and short host populations. 
 To first order, LGRBs should trace the global star formation rate (allowing for plausible biases introduced by metallicity, which could increase
 the high-z, low-Z rate). In contrast, SGRBs (assuming they have a stellar progenitor), trace the star formation rate convolved with
 a delay time distribution. In other words, in general we would expect the SGRB redshift distribution to be skewed towards lower
 redshifts, than for the LGRBs. The samples of
 \cite{LGRBs_in_bright_regions,Savaglio_2009_hosts,Svensson_2010_hostgal_sample} all measured $15-20\%$ of the optically localised LGRB host galaxies have $R>26.5$.
 However, the TOUGH survey (The optically unbiased GRB host survey) \citep{Hjorth_2012_TOUGH_catalog} find this value to be $30\%$. This difference is
 likely due to the differing redshift distributions of LGRBs between {\em Swift} and previous missions \citep{Jakobsson_2006_mean_redshift_LGRB}, with most of the non-detections
 in the TOUGH sample lying at $z>3$. 
 To these same limits $\sim 30$\% of SGRBs are ``hostless". If the underlying redshift distribution is identical between SGRBs and LGRBs 
 then we cannot draw a strong conclusion either way. However, if SGRBs are a lower redshift population on average
 then the non-detection of a similar fraction of events to the same limits would seem more surprising. 

\subsection{The kicked scenario} \label{kicks}

The possible preference for hostless SGRB sight lines to lie close to bright galaxies, with low $P_{\rm chance}$ relative to random positions on the sky, offers
 support for models in which we are observing the hostless SGRBs to arise from systems kicked from their hosts at high velocities (several hundred $\mathrm{km}\s^{-1}$),
 and potentially with significant time between their creation and explosion as a GRB ($\sim10^{9}\,\mathrm{yrs}$). For the seven secure hostless GRBs in our sample, 
 using $z=0.3$ where the redshift is unknown, the average offset is $\sim36\,\mathrm{kpc}$ with offsets ranging from $\lesssim6.6\,\mathrm{kpc}$ to $\sim85\,\mathrm{kpc}$.
 We consider the implications this model would have for hostless SGRBs here.

For the limited sample of (12) SGRBs with optical positions, confidently identified host galaxies and measured redshifts,
 the physical offsets shown in table \ref{OAG_hosts} are mostly relatively 
 low (e.g. 051221A, 050724, 070714B), with an overall mean of $\approx6\,\mathrm{kpc}$. 
 There are some, however, which are further from their host galaxy centres either 
 in the outskirts (e.g. 080905A, 071227) or outside of the host galaxy (e.g. 070429B, 090510).
 For the hostless SGRBs physical offsets $>30\,\mathrm{kpc}$ are measured for the suggested host galaxies with confirmed redshifts
 (e.g. 061201, 070809, 090515) \citep{Fong_2010_SGRBhosts_HST,Berger_2010_nohost,Rowlinson_2010_090515} and $\gtrsim6\,\mathrm{kpc}$
 for the cases with only an upper limit on the redshift (e.g. 080503) \citep{Perley_2009_080503_gals_submitted}.

Within the considerable uncertainties,
 the measured SGRB offset distribution (assuming the galaxy with the lowest  $P_{\rm chance}$ is the host in the hostless cases) 
 does appear broadly consistent with predictions for the positions of NS-NS and NS-BH binary mergers using host galaxies of mass comparable to the Milky Way
 \citep{Fong_2010_SGRBhosts_HST} and using estimated galaxy masses for the SGRBs \citep{Church_NS_offsets}. However, in some individual cases,
 the offsets from the galaxies with the lowest $P_{\rm chance}$ are surprisingly large given
 these predictions, including hostless GRBs 061201, 070809 and XRT-localised 060502B \citep{Church_NS_offsets}.

Larger offsets would also be a natural product of neutron star binary progenitors formed via a dynamical channel within globular clusters (GCs)
 where a neutron star captures a companion through three-body interactions \citep{Grindlay_2006_SGRBs_GCs}.
 For GRBs 061201 and 070809 \cite{Salvaterra_Offset_Bimodal} suggested that their bright afterglows preclude a location in the IGM and suggested the dynamical channel
 as the most likely solution. 
 However, SGRBs with optical afterglows have been detected outside their host galaxies, most likely within a low density medium if not within the IGM,
 meaning at least in some cases that the afterglow can be detected \citep{Berger_2011_SGRB_environments} (although it is not possible
 to rule out the possibility that these bursts also originated from within cluster environments). Indeed, for the hostless GRBs presented here, along with GRB\,090515 
 \citep{Rowlinson_2010_090515}, their faint afterglows could be in line with their being embedded within the IGM, with detection being possible since they are at low redshift. 

Another important consideration is that lower mass host galaxies, due to their shallower potential wells and therefore lower escape velocities,
 should typically exhibit larger burst offsets due to unbound binaries.
 For dwarf galaxies we would expect a non-negligable fraction of binary mergers to be found $>30\,\mathrm{kpc}$ from their host centres \citep{GRBoffset}.
 Using an evolving galactic potential, the merger sites may be even more diffusively distributed with respect to their host galaxies and may
 occur out to a few $\mathrm{Mpc}$ for lighter halos \citep{Zemp_NSmergeroffset}. Particularly given that for dwarf galaxies, even at moderate redshifts ($z\sim1$),
 their intrinsic faintness would make them more difficult to detect, this means that it would be more difficult to associate such a SGRB with its low mass host galaxy
 and so these cases would appear as being hostless. 
 In this case it could be that these fainter galaxies are within the halo of a larger galaxy and this is the 
 reason we're seeing a suggested association with larger galaxies at low redshift. 

\subsection{Implications for co-incident gravitational waves}
Perhaps the strongest constraints on the nature of SGRBs will come via searches for simultaneous gravitational
 wave signasl \citep{Phinney_1991_GW_DNS_merger_rate,Abadie_2010_DNS_GW_wave_signal}. For NS-NS binaries such signals should be detectable to next generation GW detectors to distances of 
 $\sim500\,\mathrm{Mpc}$ NS-NS and 
 $\sim1\,\mathrm{Gpc}$ NS-BH \citep{Abadie_2010_DNS_GW_wave_signal}, meaning that SGRBs within this horizon can have sensitive searches for
 inspirals performed.
 To date, relatively few of the detected SGRBs fall within this horizon (formally only the lowest $z=0.105$, GRB 080905A \citealt{Rowlinson_2010_080905A_Opt}
 is consistent with NS-NS or NS-BH detection),
 suggesting in common with independant analysis (e.g. \citealt{Abbott_2010_GW_GRBs}) that simultaneous detections with {\em Swift} and GW detectors will be rare.
 One possibility which could increase the event rate would be if the hostless SGRBs
 were in fact kicked from local structures within the horizon of the new advanced detectors. Our observations suggest that in general this is not the case,
 most of the candidate host galaxies are too faint to lie within this volume, and there are not many bright galaxies within several arcminutes
 (corresponding to projected distances of several hundred kpc at a distance of 100 Mpc) of the GRB positions. 

In the case of GRB 111020A there is an apparently local galaxy with low $P_{chance}$ (equal to the lowest $P_{\rm chance}$ in the field). This galaxy, at an apparent distance of $\sim 80\,\mathrm{Mpc}$ is comfortably
 within the threshold for GW detection with next generation instrumentation, although in this case the energy release of the burst of $E_{iso} \sim 10^{46}\,\mathrm{ergs}$
 would be far lower than typical for SGRBs, while the offset from the host would strongly disfavour events akin to soft-gamma repeaters 
 \citep{Hurley_2005_SGR_SGRB_origin,Tanvir_2005_SGR_SGRB_origin}. In addition, for the rest of the hostless SGRB sample we searched the NASA extragalactic
 database (NED)\footnote{http://ned.ipac.caltech.edu/} \citep{Schmitz_2011_NED} for any bright, low redshift host within $10\mathrm{'}$ of the GRB position. Galaxy 2MASX J13350593-2206302
 (also designated 6dFGS gJ133506.0-220631 in the 6dFGS catalog) is detected within $50\mathrm{''}$ of GRB\,070809 with magnitude $R=15.38$
 \citep{Jones_2004_070809_6dFGS_1,Jones_2009_070809_6dFGS_2}. With redshift $z=0.042783$ \citep{Paturel_2003_070809_bright_low-z_gal} this galaxy is
 within $190\,\mathrm{Mpc}$, again within the GW detection volume.
 These situations are rather similar to the more poorly constrained case of
 GRB\,050906 \citep{Levan_GCN_050906_bright_galaxy} whose $\gamma$-ray only position places it close on the sky to IC 328, and whose energy
 (if associated with IC 328) would be similar. None-the-less such associations on the sky are rare, of the $\sim 700$ LGRBs detected by {\em Swift} to date
 only $\sim 3$ contain bright ($R< 15$) low redshift galaxies within a few arcminutes of the burst position, that ultimately turned out not to be associated with the burst.
 This supports the  conclusions shown in Figure~9.

These results suggest that we should expect to see a handful of SGRBs within the GW horizon per year (all sky), but also imply that the hostless SGRBs likely
 contribute no more to this population than those with optically identified hosts. 

\section{Conclusions}

We have looked at the afterglow properties of three apparently hostless SGRBs: GRB\,090305A, GRB\,091109B and GRB\,111020A.
 The former, in particular, had a very faint X-ray counterpart, only identified due to the detection of the optical afterglow within the BAT error circle.
 Detection of the afterglow of GRB\,091109B in the $R$ band and GRB\,090305A in the $g^{\prime}$ band allows us to place upper redshift limits of $z\lesssim5$
 and $z\lesssim3.5$, respectively.

Deep optical observations at the GRB position after the afterglow had faded allows us to put constraints on any coincident host galaxy, specifically
 the $3\sigma$ limiting magnitude for GRB\,090305A is $r>25.69$, for GRB\,091109B the limits are $R>25.80$, $K>22.23$ and $J>21.99$ and for GRB\,111020A is $R>24.23$
 (a deeper limit of $i^{\prime}>24.4$ has also be placed by \citealt{Fong_2012_111020A_preprint_Xray}). 
 Although $r^{\prime}$ band observations make elliptical galaxies in particular difficult to detect with lower redshift limits, 
 use of a deep limit in the $K$ band would even the chances of detecting any type of host galaxy. 
 Using the deepest limiting magnitudes for GRBs 090305A and 091109B 
 we find that out to $z=1$ we uncover $\sim75\%$ of the integrated galaxy luminosity for an Sbc type galaxy, $\sim85\%$ for an Irr galaxy and $55\%$ for an Ell galaxy.
 GRB\,111020A which, even using the deepest limit available from \citealt{Fong_2012_111020A_preprint_Xray}, has a shallower limit, would uncover $\sim50\%$ (Sbc galaxy).
 It is unclear as to the status of GRB\,111020A as a hostless burst due to the presence of an unresolved 
 object $0.7\,''$ from its position which would have $P_{\rm chance}<0.01$\footnote{This case is also complicated by the relatively low Galactic latitude, which means
 that foreground stars significantly outnumber background galaxies in our images.}.
 However, for other GRBs considered here as well as no coincident host detection,
 we also find no host that can be confidently identified using $P_{\rm chance}$ values ($<1\%$). 

These GRBs represent a growing population of optically localised SGRBs with no obvious host galaxy. We have considered two possible origins for these hostless SGRBs. 
 The first is that they originate from a higher redshift, and so far unseen population of SGRBs, while the second is that they lie at lower redshift, and are kicked from
 local, and relatively bright host galaxies. 

To address these issues 
 we developed a diagnostic to assess the significance of the association of any given galaxies with a SGRB, and compared the properties
 of the sample of bursts with those of random positions on the sky. These results suggest that hostless SGRBs as a population have
 a correlation with structure at small angular scales, more so that ``average" random lines of sight. This perhaps offers evidence
 that the hostless SGRBs are in fact associated either with these bright galaxies, or with fainter galaxies associated with the same large scale structure. 
 In this case the offsets are either as a result of large scale natal kicks to the progenitors, or of their dynamical formation within globular clusters.

We note that a similar study by \cite{Berger_2010_nohost} concluded that large offsets of $15-70\,\mathrm{kpc}$ from relatively low redshift galaxies are their preferred explanation. They found
 for a high redshift solution the constraints on any underlying host galaxy implied a bimodal population of SGRBs with peaks at $z\sim0.5$ and $z\sim3$. They
 also allow for a minor contribution from NS binary mergers in globular clusters. 
 Our
 work broadly agrees with the results considered by  \cite{Berger_2010_nohost}, although critically extends it to consider the true distribution of galaxies on the sky (rather
 than average number counts), utilizing this comparison to make statistical statements on the population as a whole.  
 
Ultimately, if hostless GRBs are present at low redshift deeper observations of their locations will continue to yield null detections of their host galaxies. However, 
 for bursts associated with structure at lower redshift we may be able to ascertain if they are hosted within globular clusters via
 deep observations with either the Hubble Space Telescope (to $z<0.1$) or the James Webb Space Telescope (to $z<0.2$). Such a detection could
 offer strong evidence for the origin of SGRBs in compact binary mergers.

\section*{Acknowledgements}

We thank Edo Berger and Wen-fai Fong for discussion, and the provision
of the Gemini imaging of GRB\,110112A. RLT thanks STFC for a studentship
award. AJL, NRT, KW acknowledge receipt of STFC funding via rolling
grant awards. Based on observations made with ESO Telescopes at the
La Silla Paranal Observatory under programme IDs 088.D-0523,
084.D-0621, 082.D-0451 . Based on observations obtained at the
Gemini Observatory, which is operated by the Association of
Universities for Research in Astronomy, Inc., under a cooperative
agreement with the NSF on behalf of the Gemini partnership: the
National Science Foundation (United States), the Science and
Technology Facilities Council (United Kingdom), the National Research
Council (Canada), CONICYT (Chile), the Australian Research Council
(Australia), Ministério da Ciência, Tecnologia e Inovação (Brazil)
and Ministerio de Ciencia, Tecnología e Innovación Productiva
(Argentina). These observations are associated with programmes
GN-2009A-Q-23 and GN-2011A-Q-4. The DARK cosmology centre is funded
by the DNRF.


\bibliographystyle{mn_new}
\bibliography{091109B_paper}

\appendix

\section{SGRB sample}\label{SGRB_sample_table}

The putative host galaxies of the optically localised SGRB sample considered in this paper are detailed in table \ref{OAG_hosts}.
 This SGRB sample has been subdivided into those with small and large offsets. To do this we use the methodology of \cite{LGRBs_in_bright_regions}
 to calculate $F_{light}$ which is the fraction of total galaxy light in regions of lower surface brightness than at the position of the GRB.
 Any burst with $F_{light}>0$ must be within the host and $F_{light}=0$ indicates it is not.
 We use values of $F_{light}$ from \cite{Fong_2010_SGRBhosts_HST}, supplementing these with our own values based on our VLT imaging of GRB\,080905A and GRB\,090510.
 In the few cases when no $F_{light}$ measurement is available we assume the GRBs are on the light of their hosts, especially since these GRBs are at low offsets.  
 Where possible we have also considered a sample of SGRBs which only have XRT-localisations with the host galaxy details listed in table \ref{XAG_hosts}.

\begin{table*}
 \begin{tabular}{c c c c c c c c c}
  \hline
  GRB     & $T_{90}$$^a$ & Fluence$^a$                               & $F_{light}$ & Host mag ($r^{\prime}$(AB))$^{b}$ & z      & Offset          & Offset        & Ref \\
          &  (s)         & ($10^{-7}\mathrm{erg}\,\mathrm{cm}^{-2}$) &             &                                   &        & (arcsec)        & (kpc)         &     \\
  \hline
  \multicolumn{9}{l}{On-host GRBs}\\
  \multicolumn{9}{l}{\bf SGRBs}\\
  051221A & $1.40\pm0.20$       & $11.6\pm0.4$               & 0.54-0.65 & $21.99\pm0.09$        & 0.55   & $0.12\pm0.04$   & $0.76\pm0.25$ & [1]-[3]        \\  [1ex] 
  060121  & $1.97\pm0.06$$^c$   & $38.7\pm2.7$$^c$           &           & $26.2\pm0.3$          & $>4.0$ & $0.119\pm0.046$ & -             & [4]-[7]        \\  [1ex] 
  070429B & $0.50\pm0.10$       & $0.63\pm0.10$              &           & $23.18\pm0.10$        & 0.9023 & 0.6             & 4.7           & [8]-[10]       \\  [1ex]  
  070707  & $0.8\pm0.2$$^d$     & $2.07^{+0.06}_{-0.32}$$^d$ &           & $27.66\pm0.13$$^e$    & $<3.6$ & $0.1\pm0.3$     & -             & [11],[12]      \\  [1ex] 
  070724A & $0.40\pm0.04$       & $0.30\pm0.07$              & 0.19      & $20.56\pm0.03$        & 0.46   & $0.71\pm2.1$    & $4\pm12$      & [3],[13],[14]  \\  [1ex]  
  071227  & $1.80\pm0.00$       & $2.2\pm0.3$                & 0.19      & $20.66$$^{f}$         & 0.38   & $2.9\pm0.4$     & $15.0\pm2.2$  & [3],[15],[16]  \\  [1ex] 
  081226A & $0.4\pm0.1$         & $0.99\pm0.18$              &           & $25.79\pm0.34$        &        & $<0.5$          &               & [17],[18] \\ [1ex]
  090426  & $1.20\pm0.30$       & $1.8\pm0.3$                &           & $24.47\pm0.15$$^{g}$  & 2.6    & 0.1             & 0.8           & [19]-[21]      \\  [1ex] 
  100117A & $0.30\pm0.05$       & $0.93\pm0.13$              &           & $24.33\pm0.10$        & 0.92   & 0.6             & -             & [22],[23]      \\  [1ex]
  111117A & $0.47\pm0.09$       & $1.40\pm0.18$              &           & $23.6$   & $1.3^{+0.3}_{-0.2}$ & $1.25\pm0.20$   & $10.5\pm1.7$  & [24]-[26]      \\  
  \multicolumn{9}{l}{\bf EE GRBs}\\ 
  050709  & $0.07\pm0.01$$\,^c$ & $3.03\pm0.38$$\,^c$        &           & $21.46\pm0.2$$^{f}$   & 0.1606 & $1.33$          & $3.64$        & [7],[27]-[29]  \\  [1ex] 
  050724  & $3\pm1$             & $6.3\pm1.0$                & 0.03-0.33 & $18.36\pm0.03$$^{f}$  & 0.258  & $0.64\pm0.02$   & $2.57\pm0.08$ & [28],[30]-[32] \\  [1ex] 
  051227  & $8\pm2$             & $2.3\pm0.3$                & 0.66      & $25.78^{0.18}_{0.12}$ &        & $0.05\pm0.02$   & $<0.7$        & [16],[33],[34] \\  [1ex]  
  061006  & $130\pm10$          & $14.3\pm1.4$               & 0.56-0.63 & $24.18\pm0.09$        & 0.436  & $0.3\pm0.3$     & $3.5$         & [16],[34],[35] \\  [1ex]
  070714B & $64\pm5$            & $7.2\pm0.9$                & 0.26      & $25.39\pm0.23$$^{f}$  & 0.9225 & 0.4             & 3.1           & [10],[36]      \\  
  \hline    
  \multicolumn{9}{l}{Off-host GRBs}\\
  \multicolumn{9}{l}{\bf SGRBs} \\
  060313  & $0.70\pm0.1 $       & $11.3\pm0.5$               & 0.00-0.04 & $25.16\pm0.20$$^{e}$  &        & 0.4             & -             & [34],[37]      \\  [1ex] 
  080905A & $1.00\pm0.10$       & $1.4\pm0.2$                & 0.00      & $18.4\pm0.5$$^{f}$    & 0.122  & 9               & 18.5          & [38],[39]      \\  [1ex] 
  090510  & $0.30\pm0.10$       & $3.4\pm0.4$                & 0.00      & $23.4\pm0.07$         & 0.9    & 1.2             & 9.4           & [40]-[42]      \\ 
  \hline    
  \multicolumn{9}{l}{Hostless SGRBs}\\
  \multicolumn{9}{l}{\bf SGRBs} \\
  061201  & $0.80\pm0.10$       & $3.3\pm0.3 $               & -         & $18.09$$^{h}$         & 0.111  & $16.2$          & $32.4$        & [7],[43],[44]  \\  [1ex] 
  070809  & $1.30\pm0.10$       & $1.0\pm0.1 $               & -         & $21.8\pm0.3$$^{f}$    & 0.473  & $6.0$           & $35.4$        & [45]-[47]      \\  [1ex] 
  090305A & $0.4\pm0.1$         & $0.75\pm0.13$              & -         & $25.64\pm0.20$        &        & $1.5$           & -             & [48]           \\  [1ex] 
  090515  & $0.04\pm0.02$       & $0.2\pm0.08$               & -         & $20.2\pm0.1$          & 0.403  & $14$            & $75.2$        & [47],[49],[50] \\  [1ex]
  091109B & $0.27\pm0.05$       & $1.9\pm0.2 $               & -         & $23.88\pm0.1$$^{e}$   &        & $3.0$           & -             & [51]           \\  [1ex] 
  110112A & $0.5\pm0.1$         & $0.30\pm0.09$              & -         & $21.42\pm0.13$        &        & $19.3$          & -             & [52]           \\  [1ex] 
  111020A & $0.40\pm0.09$       & $0.65\pm0.10$              & -         & $23.336\pm0.10$       &        & $3.0$           & -             & [53],[54]      \\  
  \multicolumn{9}{l}{\bf EE GRBs}\\
  080503  & $170\pm40$          & $20\pm1$                   & -         & $27.19\pm0.2$$^{i}$   & $<4$   & $0.8$           & $<6.6$        & [47],[55],[56] \\  
  \hline
 \end{tabular}
 \caption{Host galaxy details for all well-localised SGRBs up to April 2012 (25 GRBs). The designation of on-host, off-host is described in the main text. The host galaxy listed is the one with the lowest $P_{\rm chance}$ value.
 Footnotes:
 $(a)$ $T_{90}$ and Fluence are in the $15-150\,\mathrm{keV}$ energy band unless otherwise stated.
 $(b)$ All magnitudes in this table are in the AB magnitude system and are corrected for Galactic absorption \citep{Schlegel_1998_Galactic_extinction}. Where the galaxy magnitude is unknown in the $r^{\prime}$ band a colour correction is made from the magnitude in the closest filter using one of the four standard SED templates (Sbc, Scd, Ell, Im) from \citet{Galaxy_templates} and the galaxy redshift where known. When the galaxy type is not known the Sbc template is used.
 $(c)$ \emph{HETE-2} trigger. $T_{90}$ and Fluence of GRB\,060121 are given in the $30-400\,\mathrm{keV}$ energy band.
 $(d)$ \emph{INTEGRAL} trigger. $T_{90}$ and Fluence of GRB\,070707 are given in the $20-200\,\mathrm{keV}$ energy band.
 $(e)$ For GRBs 070707, 060313 and 091109B no redshift for the most likely host galaxy is available. $R$ band magnitudes for these galaxies are converted to $r^{\prime}$ band using the Sbc template at $z=1$.
 $(f)$ $R$ band magnitude of the host galaxies of GRBs 071227, 050709, 050724, 070714B, 080905A, 070809 have been converted using the known redshift and an appropriate galaxy type.
 $(g)$ We convert the $V$ band of the host galaxy of GRB\,090426 to the $r^{\prime}$ band using the Irr galaxy template.
 $(h)$ $F606W$ band magnitude of the host galaxy of GRB\,061201 with the lowest $P_{\rm chance}$ has been converted to the $r^{\prime}$ band using known redshift and the Sbc galaxy type.
 $(i)$ $i^{\prime}$ band magnitude of the galaxy with the lowest $P_{\rm chance}$ for GRB\,110112A has been converted using $z=0$ and the Sbc galaxy template.
 $(j)$ $J$ band magnitude of the galaxy with lowest $P_{\rm chance}$ for GRB\,111020A has been converted to the $r^{\prime}$ band using $z=0$ and the Sbc galaxy type.
 $(k)$ $F606W$ band magnitude of the galaxy of GRB\,080503 with the lowest $P_{\rm chance}$ has been converted to the $r^{\prime}$ band using $z=1$ and the Sbc galaxy type.
 References: [1] \citet{Cummings_GCN_051221A_BAT_refined} [2] \citet{Soderberg_2006_051221A_AG+host} [3] \citet{Berger_2009_SGRBhosts} [4] \citet{Donaghy_2006_060121_HETE_submitted} [5] \citet{Levan_2006_060121_AG+host} [6] \citet{deUgartePostigo_2006_060121_highzhost} [7] \citet{Fong_2010_SGRBhosts_HST} [8] \citet{070429B_GCNreport} [9] \citet{Perley_GCN_070429B_host} [10] \citet{Cenko_2008_070429B_070714B_submitted} [11] \citet{McGlynn_2008_070707_INTEGRAL_analysis} [12] \citet{Piranomonte_2008_070707_Beta-OX} [13] \citet{070724A_GCNreport} [14] \citet{Berger_2009_070724A_Opt} [15] \citet{Sato_GCN_071227_BAT_refined} [16] \citet{DAvanzo_2009_3_short_opt} [17] \citet{Krimm_GCN_081226A_BAT_refined} [18] \citet{NicuesaGuelbenzu_2012_preprint_090305} [19] \citet{Sato_GCN_090426_BAT_refined} [20] \citet{Levesque_2010_090426_host} [21] \citet{Antonelli_2009_090426} [22] \citet{100117A_GCNreport} [23] \citet{Fong_2011_100117A_host} [24] \citet{111117A_GCNreport} [25] \citet{Margutti_2012_111117A_Chandra_host} [26] \citet{Cucchiara_GCN_111117A_host_Gemini} [27] \citet{Villasenor_2005_050709_analysis} [28] \citet{Troja_2008_offsets} [29] \citet{Prochaska_2006_SGRBhosts} [30] \citet{Krimm_GCN_050724_BAT_refined} [31] \citet{Prochaska_GCN_050724_redshift} [32] \citet{Berger_2005_050724_host} [33] \citet{Hullinger_GCN_051227_BAT_refined} [34] \citet{Berger_2007_highzSGRBhosts} [35] \citet{061006_GCNreport} [36] \citet{070714B_GCNreport} [37] \citet{Markwardt_GCN_060313_BAT_refined} [38] \citet{080905A_GCNreport} [39] \citet{Rowlinson_2010_080905A_Opt} [40] \citet{090510_GCNreport} [41] \citet{Rau_GCN_090510_redshift} [42] \citet{McBreen_2010_090510_LAT_analysis} [43] \citet{061201_GCNreport} [44] \citet{Stratta_2007_061201_AG} [45] \citet{070809_GCNreport} [46] \citet{Perley_GCN_070809_gals} [47] \citet{Berger_2010_nohost} [48] \citet{Krimm_GCN_090305A_BAT_refined} [49] \citet{Barthelmy_GCN_090515_BAT_refined} [50] \citet{Rowlinson_2010_090515} [51] \citet{091109B_GCNreport} [52] \citet{Barthelmy_GCN_110112A_BAT_refined} [53] \citet{Sakamoto_GCN_111020A_BAT_refined} [54] \citet{Fong_2012_111020A_preprint_Xray} [55] \citet{080503_GCNreport} [56] \citet{Perley_2009_080503_gals_submitted} } \label{OAG_hosts}
\end{table*}
\twocolumn

\begin{table*}
 \begin{tabular}{c c c c c c c c c}
  \hline
  GRB      & $T_{90}$$^a$  & Fluence$^a$                               & Host mag ($r^{\prime}$(AB))$^{b}$ & z & Offset   & Offset   & X-ray error radius & Ref \\
           &   (s)         & ($10^{-7}\mathrm{erg}\,\mathrm{cm}^{-2}$) &                                   &   & (arcsec) & (kpc)    &    (arcsec)      &     \\
  \hline
  \multicolumn{9}{l}{\bf SGRBs}\\
  050509B  & $0.13$          & $0.23\pm0.09$ & $17.18\pm0.05$$^c$ & $0.2248$     & 17.7     & 63.4     & 3.4  & [1]-[4]   \\ [1ex]
  050813   & $0.6\pm0.1$     & $1.24\pm0.46$ & $24.18\pm0.07$$^c$ & $0.719$      & 4.9      & 35.5     & 2.9  & [5],[6]   \\ [1ex]
  051210   & $1.27\pm0.05$   & $0.83\pm0.14$ & $23.80\pm0.15$     & $\gtrsim1.4$ & 2.9      & $>24.9$  & 2.9  & [7],[8]   \\ [1ex]
  060502B  & $0.09\pm0.02$   & $0.4\pm 0.05$ & $19.17\pm0.01$$^c$ & $0.287$      & 17.1     & 73.3     & 4.36 & [9],[10]  \\ [1ex]
  060801   & $0.5\pm0.1$     & $0.81\pm0.10$ & $23.20\pm0.11$     & $1.1304$     & 2.1      & 17.6     & 1.5  & [8],[11]  \\ [1ex]
  061217   & $0.212\pm0.041$ & $0.46\pm0.8$  & $23.33\pm0.07$     & $0.827$      & 1.9      & 14.4     & 1.89 & [8],[12]  \\ [1ex]
  070729   & $0.9\pm0.1$     & $1.0\pm0.2$   & $23.77\pm0.25$$^d$ &              & 10.0     & -        & 2.5  & [4],[13]  \\ [1ex]  
  090621B  & $0.14\pm0.04$   & $0.70\pm0.10$ & $21.83\pm0.11$$^d$ &              & 11.5     & -        & 5.1  & [14],[15] \\ [1ex] 
  100206A  & $0.12\pm0.03$   & $1.4\pm0.2$   & $21.3\pm0.3$$^d$   &              & 3.7      & -        & 2.1  & [16],[17] \\ [1ex] 
  100625A  & $0.33\pm0.03$   & $2.3\pm0.2$   & $\sim23$           &              & 0.5      & -        & 1.8  & [18]-[20] \\ [1ex]
  101219A  & $0.6\pm0.2$     & $4.6\pm0.3$   & $\sim24.4$$^e$      & $0.718$      & 0.5      & 3.5      & 2.4  & [21]-[24] \\ [1ex]
  101224A  & $0.2\pm0.01$    & $0.58\pm0.11$ & $20.8\pm0.2$$^f$   &              & $\sim0.5$& -        & 3.2  & [25]-[27] \\ [1ex]
  111222A  & $0.32$$^g$      & $72\pm7$$^g$  & $18.08\pm0.01$     &              & 0.48     & -        & 2.9  & [28],[29] \\ [1ex]
  \multicolumn{9}{l}{\bf EE GRBs}\\
  061210   & $85\pm5$        & $11\pm2$      & $21.00\pm0.02$     & $0.4095$     & 2.9        & 15.6     & 1.8  & [8],[30]  \\ [1ex]
  \hline           
 \end{tabular}
\caption{Host galaxy details, where available, for XRT-localised SGRBs up to April 2012 (16 GRBs). The host galaxy listed is the one with the lowest $P_{\rm chance}$ value.
Footnotes:
$(a)$ As for table \ref{OAG_hosts}, $T_{90}$ and Fluence are given in the $15-150\,\mathrm{keV}$ energy band unless otherwise stated.
$(b)$ As for table \ref{OAG_hosts} all magnitudes in this table are in the AB magnitude system and have values corrected for Galactic absorption \citep{Schlegel_1998_Galactic_extinction}. Where we do not know the galaxy magnitude in the $r^{\prime}$ band a colour correction is made from the magnitude in the closest filter using the most appropriate of the four standard SED templates (Sbc, Scd, Ell, Im) from \citet{Galaxy_templates} and the galaxy redshift where known. When the galaxy type is not known the Sbc template is used.
$(c)$ $R$ band magnitude of the host galaxies of GRBs 050509B, 050813, 060502B have been converted using the known redshift and the Ell galaxy type.
$(d)$ For GRBs 070729, 090621B and 100206A no redshift for the most likely host galaxy is available. $R$ band magnitudes for these galaxies are converted to $r^{\prime}$ band using the Sbc template at $z=1$.
$(e)$ The $i^{\prime}$ magnitude for the host of GRB 101219A has been converted to the $r^{\prime}$ band using the Sbc template at $z=0.718$.
$(f)$ $V$ band magnitude of the most likely host galaxy of GRB\,101224A has been converted using the Sbc template at $z=1$.
$(g)$ \emph{INTEGRAL} trigger. $T_{90}$ and Fluence of GRB\,111222A are given in the $20\,\mathrm{keV} - 3\,\mathrm{MeV}$ energy band and were measured by the \emph{Konus-Wind} instrument.
References: [1] \citet{Barthelmy_GCN_050509B_BAT_refined} [2] \citet{Bloom_2006_050509B} [3] \citet{Fong_2010_SGRBhosts_HST} [4] \citet{Berger_2009_SGRBhosts} [5] \citet{Sato_GCN_050813_BAT_refined} [6] \citet{Prochaska_2006_SGRBhosts} [7] \citet{Sato_GCN_051210_BAT_refined} [8] \citet{Berger_2007_highzSGRBhosts} [9] \citet{Sato_GCN_060502B_BAT_refined} [10] \citet{Bloom_2007_060502B_putative_hosts} [11] \citet{Sato_GCN_060801_BAT_refined} [12] \citet{061217_GCNreport} [13] \citet{070729_GCNreport} [14] \citet{090621B_GCNreport} [15] \citet{Galeev_GCN_090621B_opt_point} [16] \citet{100206A_GCNreport} [17] \citet{Miller_GCN_100206A_host} [18] \citet{100625A_GCNreport} [19] \citet{Levan_GCN_100625A_Gemini1} [20] \citet{Tanvir_GCN_100625A_Gemini2} [21] \citet{Krimm_GCN_101219A_BAT_refined} [22] \citet{Chornock+Berger_GCN_101219A_redshift} [23] \citet{Cenko_GCN_101219A_Gemini_host_mag} [24] \citet{Perley_GCN_101219A_Gemini_host_pos} [25] \citet{101224A_GCNreport} [26] \citet{Nugent_GCN_101224A_opt} [27] \citet{Xu_GCN_101224A_galaxy} [28] \citet{Golenetskii_GCN_111222A_Konus-Wind} [29] \citet{Siegel+Grupe_GCN_111121A_UVOT_pos} [30] \citet{061210_GCNreport} } \label{XAG_hosts}
\end{table*}

\section{Determination of $P_{detect}$} \label{Appendix_P_detect}

To perform the analysis described in section \ref{redshift_gal} we used an evolving galaxy luminosity function represented
 by a Schechter function \citep{Schechter_L_fn} where the number density of galaxies per unit luminosity, $\phi(L)$, is given by equation \ref{Schechter_fn}.

\begin{equation}
 \phi(L)dL = \phi^* \left(\frac{L}{L^*}\right)^\alpha \exp\left(-\frac{L}{L^*}\right)\frac{dL}{L^*}
\label{Schechter_fn}
\end{equation}

where $L^*$
 parameterises the position of the knee in the luminosity function. For typical values of $\alpha$,
 the bulk of the integrated luminosity is contributed by galaxies around  $L^*$. For increasing redshift we use a simple linear evolution of the luminosity function,
 investigating the magnitudes in the SDSS $r^{\prime}$ band up to $z=4$.
 We use a value of $M_r^*=-21.48$ at $z=0$ from \cite{Montero_Dorta_L-star_SDSS} and an adapted value for $M_r^*=-22.88$ at $z=2.0$
 from \cite{Ilbert_2005_LF_evolution}, using $r^{\prime}(AB)=R(AB)+0.06$ (Sbc), $=R(AB)+0.06$ (Ell) and $=R(AB)+0.04$ (Irr) appropriate for the galaxy templates used
 \citep{Galaxy_templates}.
 Similarly we also evolved $\alpha$ from $\alpha_{z=0}=-1.26$ at $z=0$ \citep{Montero_Dorta_L-star_SDSS} to $\alpha_{z=2}=-1.53$ at $z=2$ \citep{Ilbert_2005_LF_evolution}.
 The minimum galaxy luminosity detectable with our detection limits, $L_{lim}$, is determined using the \emph{python} astSED module
 as part of the astLib package.
 The probability of detection, $P_{detect}$, as function of redshift can be determined using equation \ref{P_detect}.

\begin{equation}
 P_{detect} = \left.\int_{L_{low}}^{L_{lim}} \! L\phi(L)dL \middle/ \int_{L_{low}}^{\infty} \! L\phi(L)dL \right.
\label{P_detect}
\end{equation} 

where we define $M_{low}=-10$ as the magnitude corresponding to a suitable lower limit on galaxy luminosities.
 To first order we assume that the SGRB rate is proportional to the host galaxy luminosity.
 Our chosen cosmology was $\Omega_M=0.3$, $\Omega_\lambda=0.7$ and $H_0=70\,\mathrm{km}\s^{-1}\mathrm{Mpc}^{-1}$.

\end{document}